# Thermal alteration of CM carbonaceous chondrites: mineralogical changes and metamorphic temperatures

A. J. King[1,2*], P. F. Schofield[1] and S. S. Russell[1]

[1]Planetary Materials Group, Department of Earth Sciences, Natural History Museum, Cromwell Road, London, SW7 5BD, UK.
[2]School of Physical Sciences, The Open University, Walton Hall, Milton Keynes, MK7 6AA, UK.

*corresponding author: A. J. King (a.king@nhm.ac.uk)

The CM carbonaceous chondrite meteorites provide a record of low temperature (<150°C) aqueous reactions in the early solar system. A number of CM chondrites also experienced short-lived, post-hydration thermal metamorphism at temperatures of ~200°C to >750°C. The exact conditions of thermal metamorphism and the relationship between the unheated and heated CM chondrites are not well constrained but are crucial to understanding the formation and evolution of hydrous asteroids. Here we have used position-sensitive-detector X-ray diffraction (PSD-XRD), thermogravimetric analysis (TGA) and transmission infrared (IR) spectroscopy to characterise the mineralogy and water contents of 14 heated CM and ungrouped carbonaceous chondrites. We show that heated CM chondrites underwent the same degree of aqueous alteration as the unheated CMs, however upon thermal metamorphism their mineralogy initially (300 – 500°C) changed from hydrated phyllosilicates to a dehydrated amorphous phyllosilicate phase. At higher temperatures (>500°C) we observe recrystallisation of olivine and Fe-sulphides and the formation of metal. Thermal metamorphism also caused the water contents of heated CM chondrites to decrease from ~13 wt% to ~3 wt% and a subsequent reduction in the intensity of the 3 μm feature in IR spectra. We estimate that the heated CM chondrites have lost ~15 – >65 % of the water they contained at the end of aqueous alteration. If impacts were the main cause of metamorphism, this is consistent with shock pressures of ~20 – 50 GPa. However, not all heated CM chondrites retain shock features **suggesting that some were instead heated by solar radiation. Evidence from the Hayabusa2 and ORSIRS-REx missions suggest that dehydrated materials may be common on the surfaces of primitive asteroids and our results will support upcoming analysis of samples returned from asteroids Ryugu and Bennu.**

## 1. INTRODUCTION

The CM ("Mighei-type") carbonaceous chondrite meteorites are chemically amongst the most primitive extraterrestrial materials available for study (e.g. Brearley 2006). However, the CM parent body originated beyond the water snowline (~1–3 AU) and contained ices that melted to produce fluids (Miyamoto 1991; Palguta et al. 2010; Bland and Travis 2017). This led to varying degrees of aqueous alteration that transformed anhydrous silicates and metal into a secondary mineral assemblage of abundant (>60 vol%) phyllosilicates and minor amounts (<5 vol%) of oxides, sulphides, and carbonates (Zolensky et al. 1997; Rubin et al. 2007; Howard et al. 2009, 2011). Evidence from the Mn-Cr ages of carbonate grains suggest that the CM parent body accreted to >60 km in diameter ~3.5 million years after the formation of calcium-aluminium-rich inclusions (CAIs) (Fujiya et al. 2012). The mineralogical and chemical characteristics of CM chondrites indicate that the aqueous alteration proceeded at low temperatures (<150°C) (Clayton and Mayeda 1984; Baker et al. 2002; Benedix et al. 2003; Guo and Eiler 2007; Verdier-Paoletti et al. 2017; Fujiya 2018; Vacher et al. 2019) during the first ~10 million years of the solar system (Endress et al. 1996; de Leeuw et al. 2009; Fujiya et al. 2012). As samples of a hydrous asteroid, the CM chondrites are an important tool in efforts to understand the nature, distribution, and transport of water in the protoplanetary disk.

In the global meteorite collection, there are a number of CM chondrites that following aqueous alteration also experienced thermal metamorphism. First reported in the 1980s (e.g. Skirius et al. 1986; Akai 1988; Tomeoka et al. 1989), heated CM chondrites can be identified from their unusual mineralogy, elemental and isotopic compositions, and organic chemistry relative to the unheated CM meteorites. For example, heated CM chondrites can contain an amorphous phase produced by the dehydration phyllosilicates, fine-grained, recrystallized olivine, and melted Fe-sulphide masses (Nakamura 2005; Tonui et al. 2014; Lee et al. 2016; King et al. 2019a). They are often depleted in water (Alexander et al. 2013; Garenne et al. 2014), light noble gases (e.g. $^4$He and $^{20}$Ne) (Nakamura 2006), and volatile trace elements (e.g. Cd) (Paul and Lipschutz 1990; Xiao and Lipschutz 1992; Wang and Lipschutz 1998; Lipschutz et al. 1999), while oxygen isotopic compositions can be shifted (Clayton and Mayeda 1999; Lindgren et al. 2020) and organics modified or destroyed (Kitajima et al. 2002; Quirico et al. 2018; Chan et al. 2019). By studying these different properties, and through comparison to the products of artificial heating experiments, it has been shown that some CM chondrites suffered relatively mild thermal metamorphism at temperatures of <500°C, whereas others were fully dehydrated and recrystallised at >750°C (Nakamura 2005; Tonui et al. 2014).

The mechanism, timing and duration of the metamorphic event(s) that produced heated CM chondrites are not well constrained. The metamorphism must have happened after the dominant period of hydration had ceased because phyllosilicates are usually dehydrated, although it is not clear whether the process was a single event or episodic, and tentative evidence for retrograde aqueous alteration has been reported in at least two mildly heated CM chondrites (Quirico et al. 2018; Lee et al. 2019a). It is possible that at the end of aqueous alteration, temperatures in some regions of the parent body continued to rise, leading to metamorphism of the hydrated materials. In this scenario, the heat source would have been the radiogenic decay of $^{26}$Al (half-life of 0.7 million years), and metamorphism lasted for millions of years (Miyamoto 1991; Palguta et al. 2010; Fujiya et al. 2012). However, this timescale is inconsistent with estimates from the Fe-Mg diffusion between chondrules and matrix and the structure of organics in heated CM chondrites that suggest the duration of metamorphism was much shorter, on the order of hours to several years (Nakato et al. 2008; Yabuta et al. 2010; Quirico et al. 2018).

Hypervelocity impacts were a major geological process in the early solar system and could have generated the high temperatures on short timescales required to form heated CM chondrites. Collisions into the parent body would have initially compressed and heated the target rocks, with phyllosilicates continuing to be dehydrated by residual heat after the release of impact pressure (Tyburczy et al. 1986; Tomeoka et al. 1999; Rubin 2012; Bland et al. 2014; Lindgren et al. 2015; Lunning et al. 2016). Artificial shock experiments using both terrestrial minerals and meteorites indicate that impact pressures of 15 – 30 GPa are required to initiate dehydration of phyllosilicates (Tyburczy et al. 1986; Tomeoka et al. 1999). High impact pressures would have led to the catastrophic disruption of hydrous asteroids and the formation of distinctive shock features in the target rocks (Scott et al. 1992; Rubin 2012; Lindgren et al. 2015; Michel et al. 2015; Lunning et al. 2016; Jutzi et al. 2019), however clear evidence for shock is not always present in the heated CM chondrites (e.g. Tonui et al. 2014; Lee et al. 2016). An alternative additional source of heat for near-Earth asteroids (NEAs) is solar radiation, as has been proposed for Phaethon (Ohtsuka et al. 2006; Takir et al. 2020). The surface temperature of NEAs can reach >1000°C leaving hydrous asteroids with relatively homogenous dehydrated crusts (Chaumard et al. 2012; Nakamura et al. 2015; Kitazato et al. 2021).

The visible and near-infrared (IR) reflectance spectra of heated CM chondrites closely resemble some low albedo C-type asteroids (Hiroi et al. 1993, 1996; Cloutis et al. 2012), although it has been suggested that rather than being dehydrated these bodies never experienced widespread aqueous alteration (Vernazza et al. 2015; Rivkin et al. 2019). Nevertheless, estimates of the surface water content of many C-type asteroids fall within the range of heated CM chondrites (Rivkin et al. 2003; Beck et al. 2021). These relationships suggest that materials mineralogically and compositionally similar to heated CM chondrites could be a major constituent of the regolith on primitive asteroids. This argument is further strengthened by the initial results of JAXA's Hayabusa2 and NASA's OSIRIS-REx missions, which indicate that the surfaces of the near-Earth carbonaceous asteroids Ryugu and Bennu contain mixtures of anhydrous, hydrous and dehydrated rocks (Kitazato et al. 2019, 2021; Hamilton et al. 2019; DellaGiustina et al. 2021). Quantifying the extent of aqueous and thermal alteration in the heated CM chondrites is therefore a critical step for interpreting the geological history of the samples returned from Ryugu and Bennu.

In this study we have investigated 13 CM and one C2$_{ung}$ carbonaceous chondrites reported in the literature as having experienced aqueous alteration followed by short-lived thermal metamorphism. We have characterised each meteorite using position-sensitive-detector X-ray diffraction (PSD-XRD), thermogravimetric analysis (TGA) and transmission IR spectroscopy. XRD is a sensitive indicator of the presence of dehydrated amorphous phyllosilicates, recrystallized silicates, and Fe-sulphides, and can be used to determine bulk modal mineral abundances of meteorites (Howard et al. 2009, 2011; King et al. 2015a, 2019a). Several studies have demonstrated that TGA and transmission IR spectra can further highlight mineralogical changes and quantify the abundance of water in carbonaceous chondrites (Beck et al. 2014; Garenne et al. 2014; King et al. 2015b). Crucially, each technique can be applied to the same aliquot of a meteorite, overcoming issues related to heterogeneity in what are often heavily brecciated rocks. Our aim was to combine the bulk mineralogy and water contents to evaluate the degree of aqueous alteration and thermal metamorphism in heated CM chondrites, enabling us to better understand their formation and subsequently the evolution of primitive asteroids. In addition, we compare our results to previously published chemical datasets and discuss the classification of heated CM chondrites.

# 2. HEATING STAGES AND SAMPLES

*2.1 Heating Stages*

Nakamura (2005) devised a classification scheme with four metamorphic stages that can be applied to heated CM chondrites. The metamorphic stages largely rely upon mineralogical changes identified using XRD but are also supported by petrographic observations and volatile element depletions. Fig. 1 summarises the mineralogical changes and we also briefly describe them here. In the scheme, Stage I meteorites are weakly heated, probably to temperatures of <300°C, and show no heat induced mineralogical changes in XRD patterns. Instead, the effects of thermal metamorphism are subtle and are best identified by transmission electron microscopy (TEM), Raman or IR reflectance spectroscopy. Stage II meteorites show no basal phyllosilicate reflections in XRD patterns, consistent with collapse to a partially dehydrated amorphous phyllosilicate phase as structural -OH starts to be lost during heating at temperatures of 300 – 500°C (e.g. Akai 1992; Gualtieri et al. 2012). Similarly, XRD patterns of Stage III meteorites show no phyllosilicate reflections but do have broad peaks from olivine, Fe-sulphides (troilite and pyrrhotite) and magnetite. Olivine recrystallised from fully dehydrated amorphous phyllosilicates at temperatures of 500 – 750°C produces broad diffraction peaks due to its fine-grainsize and/or low crystallinity and is distinct from primary unaltered crystalline olivine that has sharp peaks. Finally, the XRD patterns of Stage IV meteorites show only broad peaks from recrystallised olivine, pyroxene, Fe-sulphides, and metal having been heated to >750°C.

The metamorphic stages defined by Nakamura (2005) are approximate because the degree of heating depends on several factors, many of which are poorly constrained, such as the duration of heating, porosity of the affected materials, grain size and the abundance of organics. Nevertheless, we selected samples from each stage providing coverage of the full range of peak metamorphic temperatures reached by the heated CM chondrites.

*2.2 Samples*

We have studied the CM chondrites Asuka (A)-881458, Queen Alexandra Range (QUE) 93005, Elephant Moraine (EET) 87522, EET 96029, Jbilet Winselwan, Yamato (Y)-793321, Wisconsin Range (WIS) 91600 (CM$_{an}$), Y-86695, Pecora Escarpment (PCA) 91008 (CM2$_{an}$), Y-82054, Y-82098, PCA 02010, PCA 02012, and the ungrouped carbonaceous chondrite EET 83355. In the *Meteoritical Bulletin* these meteorites are all classified as "type 2" having experienced mild to moderate levels of aqueous alteration. Each meteorite has also been described in the literature as thermally metamorphosed and assigned a heating stage (Table 1) based either on its mineralogy, volatile contents, the nature of any organics present, or a combination of these datasets. For comparison we also analysed samples of the previously well characterised unheated CM2 chondrite falls Murchison and Santa Cruz.

To date there have been few systematic efforts to investigate heated CM chondrites; they have been analysed using a wide variety of techniques, often providing conflicting results, and there is little consistency between studies in how they are described and classified. Below we therefore provide a summary of the literature for each meteorite investigated in this work.

*2.3 Stage I (<300°C)*

A-881458 was first suspected as a heated CM chondrite by Akai and Tari (1997), who in TEM observations of the matrix identified an amorphous material that possibly formed through the decomposition of serpentine. Kitajima et al. (2002) reported that A-881458 contains partially graphitized carbonaceous materials, while in more brecciated regions there is evidence for dehydration and shrinkage of tochilinite-cronstedtite intergrowths (TCIs, formerly referred to in the literature as "poorly characterised phases" or PCPs) (Nakamura, 2006). However, XRD analysis of the matrix indicated the presence of unheated phyllosilicates and tochilinite, suggesting only mild thermal metamorphism at temperatures of <245°C (Nakamura 2005). Elemental characterisation of the matrix by electron microprobe (EMPA) yielded low analytical totals (~80 wt%) typical of unheated CM chondrites and there has been no significant loss of water (~20 wt%) or light (e.g. He and Ne) noble gases (Nakamura 2006).

QUE 93005 was described and classified as a highly altered CM2.1 by Rubin et al. (2007). Nearly all the chondrule silicates have been converted to phyllosilicates, while metallic Fe is rare (<0.01 vol%) and carbonates are abundant in the matrix (8 vol%) (Rubin et al. 2007). QUE 93005 is not a breccia but does have a strong petro-fabric, a common feature in the most altered CM chondrites (Rubin et al. 2007; Rubin 2012; Lindgren et al. 2015). Howard et al. (2015) reported the bulk mineralogy of QUE 93005 as 82 vol% phyllosilicate, 12 vol% anhydrous silicate (olivine and pyroxene), and 2 vol% magnetite, carbonate, and Fe-sulphides, and classified it as a CM1.3 using their petrologic scale based on phyllosilicate fraction (PSF = total phyllosilicate abundance / [total anhydrous silicate + total phyllosilicate abundance]). QUE 93005 was proposed as a weakly heated CM chondrite from the unusual Raman spectra of its insoluble organic matter (IOM), which is distinct from unprocessed primitive meteorites (Quirico et al. 2014, 2018). As Rb is more volatile than Sr, Amsellem et al. (2020) utilized the Rb-Sr system to date the timing of metamorphism in heated CM chondrites; for QUE 93005 they reported that the fractionation of Rb and Sr occurred ~1.3 Ga years ago.

EET 87522 was identified as a heated CM from its ultraviolet (UV) absorption strength and shape of the 3 μm feature – attributed to -OH/$H_2O$ bonds – in reflectance spectra (Hiroi et al. 1996). Raman spectra of the IOM in EET 87522 are consistent with weak thermal metamorphism (Quirico et al. 2018) and it contains slightly less water than unheated CM chondrites (Alexander et al. 2012, 2013; Garenne et al. 2014), although volatile trace elements (e.g. Zn, In, Bi, Tl and Cd) are not depleted (Wang and Lipschutz 1998).

EET 83355 is classified as a C2$_{ung}$ chondrite and was identified as possibly heated from the absence of the 0.7 µm feature (attributed to Fe$^{2+}$-Fe$^{3+}$ charge transfer in phyllosilicates and oxides) and the strength and shape of the UV absorption and 3 µm feature in reflectance spectra (Hiroi et al. 1997). Nakamura (2005) listed EET 83355 as a Stage I meteorite, but volatile trace elements are depleted (Xiao and Lipschutz 1992) and it contains less water (~4 wt%) than typical unheated CM chondrites (Alexander et al. 2012, 2013; Garenne et al. 2014). In addition, Quirico et al. (2018) proposed that it was strongly heated based on the Raman spectra of its IOM.

*2.4 Stage II (300–500°C)*

The petrography of EET 96029 was described in detail by Lee et al. (2016), who argued that it is one of the least altered CM chondrites (CM2.7). Many of the chondrules in EET 96029 retain glassy mesostasis and there are gehlenite-bearing CAIs. Metal is present but partially replaced by terrestrial goethite, and the Mg/Fe ratio of the matrix is low. XRD analysis indicates that the phyllosilicates were dehydrated to an amorphous material at >400°C, consistent with the replacement of tochilinite by magnetite, a loss of sulphur from the matrix, and the structure of the organics (Lee et al. 2016; Quirico et al. 2018). Different samples of EET 96029 display variable oxygen isotopic compositions and water abundances (8.2 – 11.0 wt%), suggesting that it contains lithologies that experienced a range of aqueous and/or thermal conditions (Garenne et al. 2014; Lee et al. 2016).

Jbilet Winselwan is a breccia containing diverse lithologies that underwent varying degrees of aqueous alteration (CM2.7 – ≤2.3) (King et al. 2019a). It consists of partially altered chondrules and CAIs set within a fine-grained matrix of phyllosilicates, oxides, and sulphides. Carbonates are rare and show no evidence for significant heating (Chan et al. 2017). However, some Fe-sulphides appear melted (Zolensky et al. 2016) and XRD analysis confirms that the phyllosilicates were dehydrated to an amorphous material at temperatures >400°C (King et al. 2019a). Furthermore, some samples of Jbilet Winselwan show depletions in volatile elements such as He and Cd (King et al. 2019a), whereas as others do not (Friend et al. 2018), implying that the record of thermal metamorphism is heterogeneous.

Nakamura (2006) concluded that Y-793321 is an example of dehydrated regolith material from a hydrous asteroid. It is a mildly aqueously altered, but moderately-to-highly brecciated CM chondrite, consisting of chondrules (sometimes flattened or broken down into mineral aggregates) and TCIs set within a phyllosilicate-rich matrix (Tonui et al. 2002; Nakamura 2006). EMPA of the matrix gives high analytical totals (~90 wt%) indicating low water contents, while TEM and XRD analyses suggest the presence of dehydrated amorphous phyllosilicates (Akai 1988). Olivine and pyroxene show wavy extinction and planar deformation features and carbonates are disaggregated and contain bubbles suggesting possible impact shock heating (Nakamura 2006). Harries and Langenhorst (2013) attributed Fe-sulphide-magnetite assemblages in Y-793321 to the breakdown of tochilinite at temperatures >245°C. Y-793321 contains 9.2 wt% water and is depleted in $^4$He (Nakamura 2006), although there is no evidence for any significant loss of volatile trace elements (Xiao and Lipschutz 1992; Wang and Lipschutz 1998; Lipschutz et al. 1999; Friedrich et al. 2002). In stepped combustion experiments the overall C content of Y-793321 (1.8 wt%) was found to be comparable to unheated CM chondrites, although there were slight depletions in the amount of C extracted during the lowest temperature releases (Gibson et al. 1984; McGarvie et al. 1987). In addition, Kitajima et al. (2002) reported that the carbonaceous materials in Y-793321 are partially graphitized.

WIS 91600 is a $CM_{an}$ carbonaceous chondrite (Choe et al. 2010). It is extensively brecciated and has a very fine-grained matrix that is texturally and mineralogically distinct from typical CMs, with abundant magnetite and sulphides, rare carbonates, and no tochilinite (Rubin et al. 2007; Tonui et al. 2014). Furthermore, while the bulk chemical composition of WIS 91600 is CM-like (Rubin et al. 2007), its C, N and O abundances and isotopic compositions suggest a closer affinity to either the CI chondrites or Tagish Lake (C2$_{ung}$) (Yabuta et al. 2010; Tonui et al. 2014). WIS 91600 may have originated in the outer (~10 AU) solar system (Bryson et al. 2020) but is unlike CI chondrites in having abundant chondrules that contain rare metal (~0.2 vol%) and whose phenocrysts and mesostasis have been replaced by phyllosilicates (CM2.4, Rubin et al. 2007; Choe et al. 2010). The phyllosilicates have high analytical totals (~90 wt%) and XRD and TEM analyses confirm they are a partially dehydrated amorphous material intermediate between serpentine and olivine (Tonui et al. 2014). WIS 91600 is depleted in Cd and the structure of its organics have been modified to a level that is consistent with short-duration thermal metamorphism at temperatures of 400 – 500°C (Yabuta et al. 2010). This thermal metamorphism likely led to the loss of other labile materials resulting in the IOM of WIS 91600 having a lower H/C ratio than unheated CM chondrites (Yabuta et al. 2010).

Y-86695 contains dehydrated amorphous phyllosilicates (Nakamura 2006), while its bulk oxygen isotopic composition plots just below the CM field, which Clayton and Mayeda (1999) interpreted as heating of a mildly altered CM-like precursor. Thermal metamorphism is also apparent from the shape of its 3 µm feature (Hiroi et al. 1997). Y-86695 is not a breccia and lacks a trapped solar wind component (Nakamura 2006). Nakamura (2005) lists Y-86695 as a Stage II meteorite, although carbonaceous materials are highly graphitized suggesting that its peak metamorphic temperature may have been close to ~500°C (Kitajima et al. 2002).

*2.5 Stage III (500 – 750°C)*

In PCA 91008 chondrule mesostasis is replaced by phyllosilicates and Fe-(oxy)hydroxides, although fragments of unaltered olivine are still found in the matrix (Tonui et al. 2014). Magnetite and metal occur in both chondrules and matrix, while Fe-Ni sulphides are present as porous aggregates (Tonui et al. 2014; Hanna et al. 2020). PCA 91008 is heavily weathered (grade B/C) and contains rusts and sulphates such as jarosite (Tonui et al. 2014). XRD patterns of PCA 91008 show the presence of recrystallised olivine, pyroxene, magnetite, and Fe-sulphides (Tonui et al. 2014; Hanna et al. 2020), which is supported by complex mid-IR spectra consistent with a mixture of olivine and pyroxene (Beck et al. 2014). It has a low shock stage of S1, but the matrix contains dehydration cracks and gives high analytical totals (85.6 – 99.7 wt%) indicating significant thermal metamorphism (Tonui et al. 2014; Hanna et al. 2020). TEM observations show that the matrix consists of dehydrated amorphous phyllosilicates and olivine pseudomorphs of serpentine-like materials (Tonui et al. 2014). Organics are modified (Yabuta et al. 2010; Quirico et al. 2018), depletions in Tl and Cd suggest a peak metamorphic temperature of ~500°C (Tonui et al. 2014) and based on the Rb-Sr system it was heated ~1.5 Ga years ago (Amsellem et al. 2020).

The XRD pattern of Y-82054 indicates the presence of dehydrated amorphous phyllosilicates, recrystallised olivine, and minor metal (Nakamura 2005), while carbonaceous materials are graphitic in nature, consistent with it experiencing strong thermal metamorphism (Kitajima et al. 2002). Y-82098 was identified as a highly dehydrated CM chondrite from the shape of its 3 μm feature in reflectance spectra (Hiroi et al. 1996), and a bulk oxygen isotopic composition that plots outside the CM field (Clayton and Mayeda 1999). Finally, Nakamura (2005) lists Y-82098 as a Stage III, or possible Stage IV meteorite.

*2.6 Stage IV (>750°C)*

PCA 02010 was described by Alexander et al. (2013) as appearing lighter in colour and being physically more robust than typical CM chondrites. In addition, they reported that PCA 02010 has some of the lowest H, C and N abundances and C/H and C/N ratios of any CM chondrite and suggested that it was either severely heated or is possibly a metamorphosed and/or shocked CO chondrite. The water content of PCA 02010 is low (1.1 wt%, Garenne et al. 2014), while secondary recrystallised olivine, pyroxene, Fe-sulphides, and metal were detected by XRD (Hanna et al. 2020). Transmission IR spectra lack the ~10 µm feature usually attributed to phyllosilicates in CM chondrites and are instead consistent with abundant olivine (Beck et al. 2014). Amsellem et al. (2020) reported that Rb-Sr were fractionated in PCA 02010 ~0.6 Ga years ago.

PCA 02012 is also light in colour (Alexander et al. 2013) and contains chondrules and CAIs in a fine-grained matrix that gives analytical totals of ~95 wt% (Nakato et al. 2013). TEM observations show that the matrix is predominantly made up of Fe-rich olivine and low Ca-pyroxene and often has a granoblastic texture due to crystal growth and annealing (Nakato et al. 2013). Other phases identified in the matrix include albite and porous sulphide-metal aggregates (Nakato et al. 2013). PCA 02012 has a bulk $\delta^{18}O$ value that is ~5 ‰ higher than those of typical unheated CM chondrites (Nakato et al. 2013) and is severely depleted in H, C, N, and water (1.7 wt%) (Alexander et al. 2013; Garenne et al. 2014). It is also depleted in the heavy primordial noble gases ($^{36}Ar$, $^{84}Kr$ and $^{132}Xe$) and has an $^{40}Ar/^{39}Ar$ age of ~1.6 Ga (Nakamura et al. 2015) and Rb-Sr fractionation age of ~1.2 – 1.4 Ga (Amsellem et al. 2020). Based on the maturity of the organics and limited Fe-Mg diffusion in olivine, Nakato et al. (2013) estimated that PCA 02012 was heated to >900°C in seconds to days.

## 3. EXPERIMENTAL

*3.1 PSD-XRD*

For each heated CM chondrite, we started with a ~50 mg chip free of fusion crust that was powdered (grain sizes of predominantly 5 – 15 µm) in a clean laboratory using an agate mortar and pestle. The entire powder was analysed using PSD-XRD by packing it into an aluminium sample well using the sharp edge of a spatula to produce a high degree of randomness in grain orientation and minimise the effect of preferred crystal alignments (Batchelder and Cressey 1998). XRD patterns were collected using an Enraf-Nonius PDS120 X-ray diffractometer equipped with an INEL curved 120° position-sensitive-detector (PSD) in a static geometry relative to the primary X-ray beam and sample. Cu $K_{\alpha 1}$ radiation was selected with a Ge 111 monochromator and the size of the primary beam on the sample was restricted to 0.24 mm × 2 mm using post-monochromator slits. The samples were rotated throughout the analyses, which lasted 16 hours for meteorites and 30 minutes for mineral standards.

Phase quantification utilised a profile-stripping method that has been applied to >50 carbonaceous chondrites in our laboratory (Bland et al. 2004; Howard et al. 2009, 2010, 2011, 2015; King et al. 2015a, 2017, 2019a, 2019b). The XRD pattern of a mineral standard was scaled to the same length of time as the meteorite analyses (i.e. ×32) and then reduced by a factor to match its intensity in the XRD pattern of the meteorite sample. The standard pattern was subtracted to leave a residual meteorite pattern, and this process was repeated for all phases identified in the meteorite until there were zero counts left in the residual and the sum of the fit factors was one (Fig. S1). The fit factors for the mineral standards were then corrected for relative differences in X-ray absorption to give a final volume fraction in the meteorite (Cressey and Schofield 1996). Detection limits are dependent upon both the specific phase and the bulk mineralogy of the sample and are better than 1 vol% for the meteorites in this study. Uncertainties are <0.5 vol% for crystalline phases and <1 vol% for non-crystalline phases.

*3.2 TGA*

Following the PSD-XRD measurements, ~10 – 15 mg of each meteorite powder was analysed using a TA Instruments SDT Q600 TGA. To monitor the reproducibility of the TGA measurements we also analysed five aliquots of Murchison (Table S1). The powders were loaded into an alumina crucible and mass loss from the sample recorded during heating from room temperature to 1000°C at 10°C min$^{-1}$ under a N$_2$ flow of 100 ml min$^{-1}$. The sensitivity of the TGA balance is 0.1 µg and the overall error on the measured mass loss fractions is ~0.1 % (see King et al. 2015b).

*3.3 Transmission IR Spectroscopy*

Approximately 3 mg of each meteorite powder was mixed with ~300 mg of KBr and compressed to form a 13 mm diameter pellet. Transmission mid-IR spectra (2 – 25 µm) were obtained from the pellets under a low vacuum using an offline benchtop Bruker Vertex 80V FTIR interferometer at beamline B22 at Diamond Light Source, UK. The spot size of the beam was ~15 µm and the spectral resolution was 2 cm$^{-1}$. The pellets were initially analysed having been stored in glass vials within a desiccator at room temperature; however, to minimise the effects of adsorbed terrestrial water we repeated the measurements after drying them at 300°C for two hours (Beck et al. 2014; King et al. 2015b).

**4. RESULTS**

*4.1 Bulk Mineralogy*

Fig. 2 shows representative PSD-XRD patterns for Stage I to Stage IV heated CM chondrites and Table 1 gives their bulk mineralogy. The whole profile fitting statistics (as defined by Young (1993)) for the quantitative phase analysis of the heated CM chondrites were $R_p$ = 2.2 – 4.3% and $R_{wp}$ = 3.7 – 5.6%. The main crystalline phases identified in the Stage I samples A-881458 and QUE 93005 are primary unaltered olivine and pyroxene, Fe-sulphides, magnetite, and calcite. The A-881458 pattern also contains small peaks from tochilinite, metal and gypsum. As with our previous work, for both meteorites we attribute diffraction peaks at ~12° and ~25° and broad features at ~19° and ~61° to relatively crystalline Fe-rich serpentine (cronstedtite) and fine-grained, poorly crystalline Mg-Fe serpentine, respectively (Howard et al. 2015; King et al. 2017). The PSD-XRD patterns and abundances of hydrated phyllosilicates (79 – 85 vol%), olivine (~9 vol%), pyroxene (~2 vol%), Fe-sulphides (1 – 4 vol%), magnetite (2 – 5 vol%) and calcite (~1 vol%) for A-881458 and QUE 93005 are consistent with analyses of unheated CM chondrites (Table 1, Howard et al. 2015; King et al. 2017). The abundances for QUE 93005 are in good agreement with those reported by Howard et al. (2015).

The PSD-XRD pattern of EET 87522 (representative of Stage II samples in Fig. 2) contains diffraction peaks from primary olivine (~15 vol%) and pyroxene (~7 vol%), Fe-sulphides (~1 vol%), magnetite (~2 vol%), calcite, metal, and gypsum (all <1 vol%). In contrast to A-881458 and QUE 93005, we did not observe diffraction peaks at ~12°, ~25°, ~19° and ~61° from hydrated phyllosilicates (Fig. 2). The absence of these peaks is typical for heated CM chondrites, with the loss of structural -OH causing collapse to a dehydrated amorphous phyllosilicate that does not produce coherent diffraction (Akai 1992; Nakamura 2005; Tonui et al. 2014; King et al. 2019a). However, amorphous materials still contribute weak and diffuse scattering to an XRD pattern in direct proportion to their abundance in a sample. This is further leveraged by our deliberate use of Cu $K_{\alpha 1}$ radiation, which generates a background fluorescence from all Fe-bearing phases (e.g. Bland et al. 2004).

Subsequently, during the quantitative phase analysis for EET 87522, we found that following subtraction of the crystalline phases from the diffraction pattern, we were left with a residual pattern greater than zero and the sum of the fit factors was less than one. This is consistent with the presence of an Fe-bearing amorphous component. The residual was in good agreement with the overall shape and profile intensity of our phyllosilicate standards (when scaled and excluding the diffraction peaks), although we cannot fully exclude a contribution to the diffraction pattern from primitive amorphous silicates and/or poorly crystalline rusts. We therefore conclude that EET 87522 contains partially dehydrated amorphous phyllosilicates (~75 vol%) and based on its PSD-XRD pattern is a Stage II heated CM chondrite in the scheme of Nakamura (2005).

Similarly, the PSD-XRD patterns for the other Stage II meteorites EET 96029, Jbilet Winselwan, Y-793321, WIS 91600 and Y-86695 lack the coherent diffraction peaks from hydrated phyllosilicates but do contain a significant contribution from dehydrated amorphous phyllosilicates. The PSD-XRD patterns reveal that the dehydrated amorphous phyllosilicates are abundant (64 – 79 vol%, Table 1) in these meteorites. Crystalline phases include primary olivine (5 – 23 vol%) and pyroxene (2 – 16 vol%), Fe-sulphides (2 – 4 vol%), magnetite (1 – 9 vol%), and calcite (1 – 3 vol%); a small tochilinite peak was detected for EET 96029, Jbilet Winselwan, Y-793321 and Y-86695, while several contain metal and gypsum (<1 vol%). The high magnetite abundance in WIS 91600 (~9 vol%) agrees with previous PSD-XRD analysis (Howard et al. 2015) and claims of a possible link to the CI chondrites (Yabuta et al. 2010; Tonui et al. 2014).

The Stage III meteorite Y-82054 contains dehydrated amorphous phyllosilicates (~47 vol%) but differs from the Stage II samples in having both sharp diffraction peaks from primary unaltered olivine (~20 vol%) and much broader peaks from secondary recrystallised olivine (~5 vol%) and Fe-sulphides (~6 vol%) (Fig. 2). Based on the position of the peaks, the primary olivine is Mg-rich ($Fo_{100-90}$), typical of CM chondrites, whereas the secondary olivine is Fe-bearing (~$Fo_{60-70}$) (e.g. King et al. 2019b). Pyroxene (~17 vol%), magnetite (~4 vol%), metal (<1 vol%) and gypsum (~2 vol%) are also present in Y-82054. The PSD-XRD pattern and bulk mineralogy of EET 83355 (~55 vol% dehydrated amorphous phyllosilicates, ~3 vol% secondary olivine), originally identified as a Stage I (Nakamura 2005), are more consistent with it being a Stage III meteorite. In contrast, Y-82098 contains dehydrated amorphous phyllosilicates (~77 vol%) but no secondary olivine, suggesting that it should be classified as a Stage II heated CM chondrite (Table 1).

For PCA 91008 we identified primary (~23 vol%) and secondary (~54 vol%) olivine, pyroxene (~17 vol%), Fe-sulphides (~1 vol%), magnetite (~3 vol%), plus gypsum and rusts (likely terrestrial weathering products). However, there was no evidence for dehydrated amorphous phyllosilicates, suggesting that it had fully dehydrated and recrystallised back into anhydrous silicates, and hence PCA 91008 is a Stage IV heated CM chondrite. The other Stage IV samples, PCA 02010 and PCA 02012, consist of primary (10 – 28 vol%) and secondary (39 – 66 vol%) olivine, pyroxene (~16 vol%), Fe-sulphides (4 – 13 vol%), metal (~1 vol%), and unlike the other samples analysed in this study ~3 vol% feldspar. No significant dehydrated amorphous phyllosilicates were observed in the Stage IV meteorites.

*4.2 TGA*

Fig. 3 shows examples of the mass loss (wt%) and derivative (wt%/°C) curves for Stage I to Stage IV heated CM chondrites. The TGA results for all the samples investigated in this work are summarised in Table 2. The derivative curve shows how the rate of mass loss changes during heating, with peak positions (i.e. mass loss events) as a function of temperature characteristic of mineral groups and/or individual phases. Previous studies of aqueously altered CM and CI chondrites have demonstrated that terrestrial adsorbed water is removed at temperatures of <200°C (with sulphates breaking down at ~120°C), -OH/$H_2O$ from Fe-(oxy)hydroxides is released between 200 – 400°C, dehydration and dehydroxylation of phyllosilicates occurs at 400 – 770°C, and the breakdown of carbonates and release of $CO_2$ takes place at 770 – 900°C (Garenne et al. 2014; King et al. 2015a).

All of the meteorites (both falls and finds) in this study show mass loss (1.0 – 8.6 wt%) at <200°C that we attribute to terrestrial alteration, i.e. the removal of adsorbed water at ~50°C and the breakdown of sulphate minerals at ~120°C, and this temperature range is not considered further. Mass loss between 200 – 400°C due to dehydration of Fe-(oxy)hydroxides, which may be a product of parent body aqueous alteration and/or terrestrial weathering, ranges from 0.5 wt% to 4.1 wt%.

For the unheated CM chondrites Murchison and Santa Cruz, the largest mass loss events (~7 wt%) occur between 400 – 770°C, the temperature regime where -OH/$H_2O$ is released from serpentine minerals (Garenne et al. 2014; King et al. 2015b). The Stage I meteorite EET 87522 shows comparable mass loss (6.2 wt%) at these temperatures, while A-881458 and QUE 93005 lose 9.8 wt% and 11.2 wt%, respectively. Similarly, the most significant mass loss for the Stage II meteorites also occurs between 400 – 770°C, with values ranging from 5.6 wt% (Y-793321) to 8.7 wt% (EET 96029). Mass loss for the Stage III meteorites Y-82054 (4.3 wt%) and EET 83355 (1.9 wt%) falls below this range, while the Stage IV meteorites, with the exception of PCA 91008 (5.1 wt%), show only minor mass loss (1.5 – 2.1 wt%) between 400 – 770°C. Small mass loss events (0.2 – 2.0 wt%) were observed for all of the meteorites from 770 – 900°C due to the breakdown of carbonates.

Several of our samples (Murchison, EET 87522, EET 96029, WIS 91600, EET 83355, PCA 02010 and PCA 02012) were also analysed using TGA by Garenne et al. (2014), and in general we find good agreement between the datasets. The most notable discrepancy is for EET 96029, with our aliquot losing more mass (8.6 wt%) between 400 – 770°C than the Garenne et al. (2014) sample (4.5 wt%), which is likely due to the heterogeneity of this meteorite (Lee et al. 2016).

Garenne et al. (2014) proposed that TGA can be used to estimate the water abundance of CM chondrites by assuming that mass loss between 200 and 770°C is entirely due to dehydration and dehydroxylation of Fe-(oxy)hydroxides and phyllosilicates. A limitation of this method is that additional mass loss in this temperature range could also result from the release of $SO_2$ as Fe-sulphides breakdown, and $H_2O$, $CO_2$ and $SO_2$ from the decomposition of the refractory IOM present in CM chondrites at ~1–2 wt% (Alexander et al. 2007). Nevertheless, TGA-derived water abundances have been shown to correlate well with those measured in the same meteorites using other analytical methods (Garenne et al. 2014; King et al. 2015b). From our TGA data we determine water abundances of ~11 wt% in the unheated CM chondrites, and ~13 wt% in the Stage I, 7.5 – 11.0 wt% in the Stage II, 3.4 – 6.3 wt% in the Stage III, and 2.9 – 7.4 wt% in the Stage IV samples (Table 2).

*4.3 Transmission IR Spectra*

Fig. 4a shows examples of transmission mid-IR (5 – 25 µm) spectra for Stage I to Stage IV heated CM chondrites. The main features in the spectra of the Stage I samples A-881458 and QUE 93005 are a narrow peak at ~9.9 µm, and broader peaks at ~16 µm and ~22 µm. They can be attributed to the abundant phyllosilicates present in these meteorites; features at ~9.9 µm and ~22 µm are due to Si-O stretching and bending modes, and the feature at ~16 µm may be related to -OH. In addition, for the Stage I meteorites there is a subtle shoulder on the ~9.9 µm peak at ~11.2 µm, consistent with the presence of small amounts of olivine (Fig. S2). Other minor features occur at ~6 – 7 µm and ~8.5 µm and are related to carbonates and sulphates, respectively. Overall, we find that the mid-IR spectra of A-881458 and QUE 93005 are very similar to Murchison and Santa Cruz and previous measurements of other unheated CM chondrites in transmission (Beck et al. 2014).

Mid-IR spectra for the Stage II meteorites also show a narrow peak at ~9.9 µm and a broad feature at ~22 µm but, with the exception of WIS 91600 (Fig. S3), the ~16 µm feature is much less prominent relative to that of the Stage Is (Fig. 4a). The Stage II meteorites have distinct shoulders at ~11.2 µm from olivine and most show carbonate peaks at ~6 – 7 µm, while EET 87522 and EET 96029 have a feature at ~8.5 µm from sulphates. The mid-IR spectra for EET 96029 and WIS 91600 are in good agreement with those reported for these meteorites by Beck et al. (2014). The spectrum for Jbilet Winselwan differs from other CM chondrites as it contains multiple intense peaks (at ~8.5 µm, ~8.9 µm, ~14.8 µm and ~16.8 µm) from sulphates (Fig. S3, King et al. 2019a).

The mid-IR spectra of the Stage III meteorites Y-82054 and EET 83355 are more complex, with a series of features of comparable intensity at ~9.3 µm, ~9.9 µm and ~11.2 µm, and less intense features at ~13.5 µm, ~14.5 µm, ~20 µm and ~24 µm (Fig. 4a). The ~9.9 µm peak is likely due to dehydrated phyllosilicates, and through comparison to standards we attribute the ~11.2 µm feature to olivine, the ~9.3 µm, ~13.5 µm and ~14.5 µm features to pyroxene, and the ~20 µm and ~24 µm to a combination of all three phases (Fig. S2). In contrast, the mid-IR spectra of the Stage IV samples PCA 02010 and PCA 02012 are dominated by the signature of olivine, with features at ~10.2 µm, ~11.2 µm, ~17 µm, ~19 µm and ~25 µm, although each also has a shoulder at ~9.3 µm from pyroxene (Fig. 4a). PCA 91008 is slightly different, with less prominent features at ~17 µm, ~19 µm and ~25 µm, and an intense peak at ~8.5 µm from sulphates (Fig. S3). A similar spectrum for PCA 91008 was presented by Beck et al. (2014).

Fig. 4b shows examples of the "3 µm" region (2.6 – 3.3 µm) in transmission IR spectra (collected after drying at 300°C) of the Stage I to Stage IV heated CM chondrites. In unheated CM chondrites the 3 µm region consists of a broad absorption feature extending from 2.7 to 3.2 µm that is attributed to a combination of -OH/$H_2O$ within hydrous phases such as phyllosilicates, tochilinite, sulphates and rusts (Sato et al. 1997; Osawa et al. 2005; Beck et al. 2010, 2014; Takir et al. 2013). Often the feature is asymmetric, with a sharper peak at ~2.72 µm from the -OH stretching mode that is assigned to hydroxyl ions structurally bound within the minerals. This spectral region is susceptible to terrestrial alteration but drying the pellets at 300°C prior to analysis has been shown to minimise the effects in CM chondrites (Beck et al. 2010, 2014; Takir et al. 2013; King et al. 2015b; Bates et al. 2020). However, 300°C is above the breakdown temperature of tochilinite, sulphates and rusts; modification of the spectra may have occurred during the drying but as these are minor phases any effects were likely minimal. The 3 µm regions of the Stage I meteorites A-881458 and QUE 93005 are like those of unheated CM chondrites, whereas 3 µm features for the Stage IIs are similar in shape but, at least qualitatively, usually appear to be less intense and are shifted to higher wavelengths. This trend continues for the Stage III and Stage IV meteorites, which do have 3 µm features but with a much lower intensity.

Beck et al. (2014) developed two criteria to quantify the level of hydration in CM chondrites from transmission mid-IR spectra. The first is the integrated intensity of the 3 µm region (from 2.65 – 2.85 µm) normalised (to account for any variability in pellet thickness) to that of the $SiO_4$ tetrahedron feature (7.7 – 14.3 µm), while the second infers the abundance of olivine relative to the phyllosilicates (integrated intensity from 11.15 – 11.25 µm / 10.94 – 10.98 µm). Here, we have applied a modified approach to quantify the intensity of the 3 µm/$H_2O$ (2.68 – 2.85 µm / 9 – 12 µm) and 11 µm/olivine (11.2 – 11.4 µm / 9.8 – 10.1 µm) features. The calculated intensities are given in Table 2 and Fig. S4 confirms that there is a good correlation between both the 3 µm feature and TGA derived water contents and the 11 µm feature and olivine abundances from XRD.

## 5. DISCUSSION

Heated CM chondrites are highly complex rocks where the record of aqueous alteration has been overprinted by thermal metamorphism. The extent of aqueous and thermal alteration in these meteorites is not well constrained, while there also remains some doubt as to how they are related to the unheated CM chondrites. Disentangling the effects of both stages of alteration is a challenging but crucial step towards understanding the evolution of primitive asteroids. In the following, we evaluate the degree of aqueous alteration and peak metamorphic temperatures for the heated CM chondrites and review the potential heat sources. We then describe the relationship between aqueous and thermal alteration. Finally, we conclude by comparing our results to published chemical datasets and discussing the classification of heated CM chondrites.

*5.1 Degree of Aqueous Alteration*

The CM chondrites span a petrologic range, from mildly aqueously altered type 2 meteorites to fully hydrated type 1s (Zolensky et al. 1997; Rubin et al. 2007; Howard et al. 2015). The extent of aqueous alteration in CM chondrites reflects the physical and chemical conditions (e.g. water/rock ratio, pH, temperature etc) on hydrous asteroids in the early solar system. With increasing aqueous alteration, the abundance of phyllosilicates and water in the rocks increased, and the abundance of olivine and pyroxene decreased. Subsequently, these properties are now regularly used to infer the relative degree of aqueous alteration in CM chondrites (Alexander et al. 2013; Beck et al. 2014; Garenne et al. 2014; Howard et al. 2015). These classification schemes are not directly applicable to heated CM chondrites because the effects of aqueous alteration were modified by thermal metamorphism; as we have shown, heated CMs typically have lower water contents and potentially higher olivine abundances than their unheated counterparts. Present-day water and olivine abundances may therefore help to identify heated CM chondrites, however they cannot be used to assess the extent of aqueous alteration prior to thermal metamorphism. For example, in Alexander et al. (2013) heated CMs are assigned a petrologic type based on their bulk hydrogen abundance, but this does not consider the additional complexity of hydrogen loss during the metamorphic event.

The bulk mineralogy of a heated CM chondrite as determined from PSD-XRD can be used to evaluate its degree of aqueous alteration prior to thermal metamorphism. Howard et al. (2015) demonstrated that the PSF is an excellent proxy for the degree of alteration. In this scheme, the PSF is converted to a petrologic type from 3.0 (unaltered) to 1.0 (complete alteration), with subtypes defined by 5 vol% increments in phyllosilicate abundance. The petrologic types of the heated CM chondrites in this study were calculated using this method and are given in Table 1. Within the uncertainties, both Stage I meteorites A-881458 and QUE 93005, where we still observe coherent diffraction from the phyllosilicates, are petrologic type 1.3-1.2.

The PSF classification can be applied to Stage II meteorites by assuming that the amorphous material in these samples represents phyllosilicates dehydrated by thermal metamorphism. This assumption is reasonable as phyllosilicates are known to produce an amorphous material upon heating (e.g. Akai 1992; Gualtieri et al. 2012; Lindgren et al. 2020) and petrographic evidence supports their presence in the Stage II meteorites. Primary amorphous silicates and rusts could occur, but their contribution to the bulk mineralogy is relatively minor. For the Stage II meteorites we determine petrologic types prior to thermal metamorphism of between 1.7 (EET 96029 and Y-793321) to 1.2 (WIS 91600), values that cover the same range as those previously reported for unheated CM chondrites (~1.7 – 1.1, Howard et al. 2015; King et al. 2017; Lee et al. 2016).

The PSD-XRD and IR analyses show that the Stage III meteorites Y-82054 and EET 83355 consist of both dehydrated amorphous phyllosilicates and secondary olivine recrystallised during thermal metamorphism. The recrystallisation of secondary olivine from dehydrated amorphous phyllosilicates starts at ~500°C (e.g. Akai 1992; Gualtieri et al. 2012). Summing together the present-day abundance of the dehydrated amorphous phyllosilicates and secondary olivine enables us to estimate the phyllosilicate abundance at the end of aqueous alteration in the Stage III meteorites. After correcting for a decrease in volume of ~50% during the transformation of phyllosilicate back into olivine (Deer et al. 1982), this results in a petrologic type of 1.9-1.8 for Y-82054 and 1.8-1.7 for EET 83355. Similarly, the Stage IV samples are predominantly secondary olivine; if we assume the secondary olivine was originally a phyllosilicate at the point of thermal metamorphism this results in petrologic types of between 2.1 (PCA 02012) to 1.6 (PCA 02010) for the Stage IV meteorites. Heating experiments suggest that secondary pyroxenes can also recrystallise from dehydrated amorphous phyllosilicates at high temperatures (>700°C, Akai 1992), but we are unable to unambiguously resolve secondary and primary pyroxene in the PSD-XRD patterns and the IR spectra confirm that olivine is the major silicate phase in the Stage IV meteorites.

*5.2 Peak Metamorphic Temperatures*

Constraining the peak metamorphic temperatures experienced by CM chondrites is important for understanding the sources of heat and overall thermal evolution of the parent body. The main methods for estimating the peak metamorphic temperatures of heated CM chondrites are thermally induced mineralogical changes (e.g. breakdown of phyllosilicates, recrystallisation of olivine etc) identified from XRD patterns, and through comparison to artificial heating experiments (Akai 1992; Nakamura 2005; Tonui et al. 2014; Chan et al. 2019; King et al. 2019a; Lindgren et al. 2020). Nakamura (2005) used XRD, plus supporting information from TEM studies, to define four heating stages for the CM chondrites (see section 2.1). The temperature ranges for each heating stage are relatively broad (~200 – 300°C) as the mineralogical changes are related to factors such as composition, porosity, grainsize and duration of heating. More specific temperatures for individual meteorites have been inferred from characteristics such as major, volatile and trace element abundances and isotopic compositions, IR and Raman spectroscopy (Paul and Lipschutz 1990; Xiao and Lipschutz 1992; Hiroi et al. 1993, 1996; Wang and Lipschutz 1998; Lipschutz et al. 1999; Kitajima et al. 2002; Nakato et al. 2008; Choe et al. 2010; Yabuta et al. 2010; Tonui et al. 2014; Mahan et al. 2018; Quirico et al. 2018; King et al. 2019a); however, there is often little agreement in the estimated temperatures as the techniques measure different properties (e.g. minerals, organics etc) and the meteorites are fine-grained, heterogenous rocks that are challenging to analyse. This is apparent in our study, where we suggest that EET 87522 and Y-82098 are Stage II samples, EET 83355 is a Stage III, and PCA 91008 is a Stage IV based on their XRD patterns and bulk mineralogy. Of these meteorites, an XRD pattern has only previously been reported for PCA 91008, which confirmed the presence of abundant secondary olivine (Tonui et al. 2014), whereas metamorphic stages for the other meteorites were defined using other analytical methods.

The bulk water abundance is a helpful criterion for identifying heated CMs (although petrographic observations are also required to confirm the CM classification) but cannot be used alone to estimate peak metamorphic temperatures as it is also related to the degree of alteration prior to metamorphism. For example, unheated CM chondrites that underwent only mild aqueous alteration have water contents comparable to some heated CMs. However, our combined XRD, TGA and IR dataset allows us to determine the relative degree of both aqueous and thermal alteration for the heated CM chondrites.

Fig. 5 shows phyllosilicate abundances (both hydrated and dehydrated) plotted against TGA derived water abundances for unheated and heated CM chondrites. This figure plotted using the intensity of the 3 µm feature rather than TGA derived water abundance shows the same overall trends (Fig. S5). For unheated CM chondrites, with more extensive aqueous alteration there is an increase in the abundance of phyllosilicates and water that they contain. Thermal metamorphism at temperatures between 300 – 500°C (Stage II) causes a loss of -OH/$H_2O$, moving samples vertically downwards from this trend but with no change in phyllosilicate abundance. At temperatures >500°C (Stage III), -OH/$H_2O$ continues to be lost and the phyllosilicate abundance decreases as they are converted back into anhydrous silicates. At temperatures >750°C (Stage IV), samples would be fully dehydrated, olivine-rich rocks, plotting towards the bottom left corner of Fig. 5.

Fig. 5 shows that the Stage I samples A-881458 and QUE 93005 have water contents consistent with their phyllosilicate abundances; they show no signs of dehydration, which is also supported by XRD patterns with coherent diffraction peaks from hydrated phyllosilicates. We are unable to distinguish A-881458 and QUE 93005 from the unheated CM chondrites using our dataset and so peak metamorphic temperatures must have been lower than the breakdown temperature of serpentine (~300–400°C) (Akai 1992; Lindgren et al. 2020). Evidence for low level thermal metamorphism in A-881458 includes a possible decomposed serpentine phase in the matrix (Akai and Tari 1997) and partial shrinkage of TCIs, although EMPA totals and light noble gas abundances are typical of unheated CM chondrites (Nakamura 2006). QUE 93005 has a strong petro-fabric (Rubin et al. 2007; Rubin 2012) and was identified as weakly heated from Raman analyses of its IOM (Quirico et al. 2014, 2018). It is likely that organics are more sensitive to low peak metamorphic temperatures than the silicate mineralogy probed by our methods, although we cannot rule out the possibility that thermal metamorphism was overprinted by a later period of aqueous alteration. Quirico et al. (2018) suggested that QUE 93005 experienced retrograde aqueous alteration, and Lee et al. (2019a) proposed a similar scenario for the CM chondrite MET 01075.

The XRD patterns of the Stage II meteorites all indicate the presence of dehydrated amorphous phyllosilicates and Fig. 5 shows that in general, with the exception of EET 96029 and Jbilet Winselwan, they are clearly depleted in water relative to unheated CM chondrites. Using the correlation shown on Fig. 5 between phyllosilicate and water abundances in the unheated CMs, we can estimate the amount of water in the Stage II meteorites at the onset of thermal metamorphism and subsequently their level of dehydration. From this we calculate that EET 96029 has experienced no significant loss of water, whereas Jbilet Winselwan has lost 13 (±3) % of its original water, Y-86695 (26 ± 6), Y-793321 (26 ± 6) and EET 87522 (29 ± 7) between ~25 – 30 %, and WIS 91600 (34 ± 8) and Y-82098 (34 ± 8) ~35 %. The TGA derived

water abundance (8.2 wt%) for EET 96029 given by Garenne et al. (2014) indicates a water loss of ~20 %, but that was from a separate sample to our study, in which we have collected XRD and TGA data from exactly the same aliquot of the meteorite. We are unable to determine specific peak temperatures for the Stage II samples but if we assume that the duration of heating was the same for each meteorite then the relative degree of metamorphism is EET 96029 < Jbilet Winselwan < Y-86695 = Y-793321 = EET 87522 < WIS 91600 = Y-82098.

The situation for the Stage III meteorites is more complex; there are only two samples in this study that fall into this temperature range, both of which contain less water than the unheated and Stage I and II CM chondrites (Fig. 5). The dehydrated amorphous phyllosilicate abundances are also lower in these meteorites due to recrystallisation back into secondary olivine at >500°C. Summing together the abundance of dehydrated amorphous phyllosilicate and secondary olivine as a proxy for the phyllosilicate abundance at the start of thermal metamorphism indicates that EET 83355 has lost 64 (±16) % of its original water and Y-82054 has lost 27 (±7) %. This suggests that EET 83355 experienced a higher peak metamorphic temperature than Y-82054, and that Y-82054 is maybe intermediate between a Stage II/III.

Using the same approach for the Stage IV samples, which is complicated by the possible formation of secondary pyroxene (>700°C, Akai 1992) that cannot be easily resolved in the PSD-XRD patterns, we estimate that PCA 02010 and PCA 02012 have lost 64 (±15) % and 59 (±18) of their original water, respectively, and PCA 91008 has lost only 17 (±4) %. Previous studies have shown that PCA 02010, PCA 02012, and especially PCA 91008, are highly weathered and contain terrestrial rusts and sulphates (Tonui et al. 2014; Hanna et al. 2020). Rusts such as goethite dehydrate at temperatures of 200 – 400°C, while jarosite, which has previously been identified in PCA 91008 (Tonui et al. 2014), has mass loss events at ~400°C and ~650°C. Terrestrial weathering products can therefore contribute to the TGA derived water abundances, which for the meteorites are calculated from the mass loss between 200 – 770°C (Fig. 3, Garenne et al. 2014; King et al. 2015b). Consequently, the TGA water abundances for PCA 91008, and also PCA 02010 and PCA 02012, may be partly enhanced by terrestrial contamination. However, from the XRD we find that the mineralogy of these meteorites is dominated by olivine, consistent with Stage IV meteorites and previous reports (Nakato et al. 2013; Tonui et al. 2014), and that their transmission mid-IR spectra are distinct from the Stage III meteorites (Fig. 4a).

*5.3 Heat Sources*

Our data show that the heated CM chondrites experienced varying degrees of aqueous alteration (petrologic types 2.1 to 1.2) and a wide range of peak metamorphic temperatures (~200°C to >750°C). Thermal metamorphism was short-lived, with the most likely mechanisms capable of generating these temperatures on this timescale being impacts and solar radiation (Nakato et al. 2008; Yabuta et al. 2010; Quirico et al. 2018). However, it is unclear whether the heated CM chondrites record a single metamorphic event or if multiple impacts and/or radiation are responsible for their overall level of dehydration. Below we review the evidence and implications for impact and radiative heating of the CM parent body.

The degree of impact shock would have been strongly influenced by the size, mass, density, composition, local geology, and relative velocity of both the impactor and parent body, resulting in highly variable peak metamorphic temperatures. Wakita and Genda (2019) showed that for impacts at relative velocities typical for the main asteroid belt (5 $kms^{-1}$), hydrous materials can avoid significant heating, perhaps explaining why there are relatively few heated CM chondrites in the global meteorite collection. Much higher impact pressures >15 GPa are needed to dehydrate phyllosilicates (Tyburczy et al. 1986; Tomeoka et al. 1999). Sekine et al. (2015) reported water loss in Mg-rich serpentine as a function of shock pressure, and if impacts were the primary heat source, then based on that relationship we find that water depletions for the heated CM chondrites in this study (~15 – >65 %) indicate shock pressures of 20 – 50 GPa.

The largest collisions in the early solar system caused catastrophic disruption of hydrous bodies, creating asteroid families of ~$10^4 – 10^5$ smaller objects and re-accumulated rubble-piles (Scott et al. 1992; Michel et al. 2015; Jutzi et al. 2019). Michel et al. (2020) recently demonstrated that rubble-pile asteroids formed during the break-up of a 100 km diameter body would have different levels of hydration, as is observed for Ryugu and Bennu, due to variations in the peak metamorphic temperatures experienced by the disrupted materials. Impact heating is consistent with the range of temperatures recorded in the heated CM chondrites, including reports of heterogenous metamorphism (Zolensky et al. 2014; Lee et al. 2016; King et al. 2019a), and is also supported by the presence of shock features in meteorites such as Y-793321 (Nakamura 2006).

Impacts are stochastic events and could happen at any point in an asteroid's history. For rubble-piles, impacts continued after the catastrophic disruption of the original body, with the largest craters of Bennu formed during a 0.1 – 1 billion-year residence in the inner main belt prior to its transport into the near-Earth population (Michel et al. 2020). Nevertheless, bulk Ar-Ar and Rb-Sr fractionation ages for several heated CM chondrites indicate that thermal metamorphism took place at least ~3 billion years after CAIs and are remarkably similar to the estimated ages for the formation of several C-type asteroid families in the main belt (Nakamura et al. 2015; Amsellem et al. 2020).

Shock features are not found in all heated CM chondrites (e.g. Y-86695, PCA 91008, EET 96029), implying that some were instead metamorphosed by solar radiation (Nakamura 2006; Tonui et al. 2014; Lee et al. 2016). Asteroids in the main belt are not significantly affected by radiative heating, whereas the surface temperatures of NEAs can reach ~1500°C at 0.05 AU and ~300°C at 0.5 AU, with temperatures decreasing rapidly with depth to leave dehydrated crusts (Chaumard et al. 2012; Nakamura et al. 2015; Kitazato et al. 2021). The dynamical lifetime of NEAs is on the order of ~10 million years (Gladman et al. 2000) suggesting that any solar heating would be a recent process. If solar radiation was the main cause of the metamorphism, then the heterogenous effects observed in meteorites such as EET 96029 could be explained by impact-induced mixing of regolith materials exposed to different intensities of radiative heating. Further evidence for radiative heating comes from depletions of light cosmogenic gases in some heated CM chondrites that could result from spending a period of time close to the Sun (Nakamura 2005; Nakamura et al. 2015).

Impacts and solar radiation both probably played a role in the post-hydration thermal metamorphism of the CM parent body. Our bulk data enable us to quantify the extent of aqueous and thermal alteration for heated CM chondrites; however, to unambiguously resolve between impacts and solar radiation requires identification and characterisation of shock features. Nearly all CM chondrites are breccias but most apparently experienced only low shock pressures (<5 GPa) and are assigned S1 shock stages, partly because alteration has often erased any olivine and feldspar microstructures (Scott et al. 1992; Rubin 2012; Lindgren et al. 2015; Lunning et al. 2016). We suggest that future petrographic studies systematically search for other indicators of shock, including petro-fabrics, crushed and broken chondrules and carbonates, and comminuted matrix, in the heated CM chondrites.

*5.4 Relationship between Aqueous Alteration and Thermal Metamorphism*

Models suggest that due to radiogenic decay, the interior of a primitive hydrous asteroid would have been warmer than the exterior, potentially resulting in a layered body with materials progressively more altered with depth (e.g. Palguta et al. 2010). However, the situation in the early solar system was more complicated, with impacts providing additional heat to melt ices. Radiative heating was not a factor at this stage because the CM parent body formed at >1 AU. There is a good correlation between the degree of aqueous alteration and strength of petro-fabrics and particle alignment in unheated CM chondrites (Rubin 2012; Lindgren et al. 2015). Rubin (2012) suggested that impacts into the parent body prior to the main period of hydration produced strong petro-fabrics and extensive fractures in the target rocks that facilitated fluid mobilisation and higher water/rock ratios, causing more aqueous alteration in these regions. Lindgren et al. (2015) reported that fractures in some CM chondrites must have been opened when fluids were already present, whereas other samples only record impacts post-aqueous alteration. Furthermore, Hanna et al. (2015) argued that in Murchison multiple episodes of fracturing occurred, and that some aqueous alteration post-dated or was contemporaneous with the deformation processes.

The metamorphic event(s) that produced heated CM chondrites occurred after the main period of aqueous alteration on the CM parent body. As impacts and solar radiation are predominantly surface processes, if the parent body was layered at the point of metamorphism, then it might be expected that the least aqueously altered CM chondrites would dominate the target rocks and therefore witness the highest peak metamorphic temperatures and greatest levels of dehydration. However, there is evidence from U isotopes that some CM chondrites experienced fluid flow within the past million years, possibly due to impact or radiative heating of fragments ejected from the parent body to temperatures sufficient to melt any remaining ices (Turner et al. 2021). This requires the CM parent body to have retained ices over the entire lifetime of the solar system and implies that recent heating from impacts and/or solar radiation could lead to not only dehydration, but also hydration of the target rocks. This may be the case for the Stage I meteorites A-881458 and QUE 93005, where we find that low levels of metamorphism have been overprinted by aqueous alteration. In contrast, we have shown that phyllosilicates in the Stage II–IV meteorites are dehydrated and recrystallised indicating that thermal metamorphism was the last major event they experienced.

Fig. 6a shows a possible correlation between the degree of aqueous alteration and peak metamorphic temperatures of the heated CM chondrites. At first glance it appears that the least altered meteorites experienced the highest temperatures, consistent with impact or radiative heating of a layered parent body. As discussed, determining the degree of aqueous alteration in the Stage IV meteorites is challenging as they may contain pyroxenes recrystallized from the phyllosilicates that we cannot easily detect using PSD-XRD (see section 5.1). The PSF values of the Stage IV samples are therefore lower limits, although relatively low degrees of aqueous alteration for PCA 91008 and PCA 02012 are supported by petrographic studies that reported unaltered chondrules in these meteorites (Nakato et al. 2013; Tonui et al. 2014). However, closer inspection of the data for just the Stage II meteorites shows the opposite trend, with the most altered samples seemingly having lost a greater amount of water due to higher metamorphic temperatures (Fig. 6b, see section 5.2). This is supported by the data of Quirico et al. (2018), who found that the most aqueously altered CM chondrites contain the most thermally processed IOM. Therefore, our limited dataset precludes confirmation of a direct relationship between the degree of aqueous alteration and peak metamorphic temperatures. This observation is intriguing nonetheless, a lack of an obvious correlation is perhaps the most expected outcome because even low impact pressures can induce mixing that leads to asteroid surfaces containing mixtures of variably altered materials (Scott et al. 1992; Rubin 2012; Bland et al. 2014; Lunning et al. 2016; Michel et al. 2020), while the meteorites also potentially sample multiple parent bodies that formed at different times and places in the solar system.

*5.5 Chemical Properties of Heated CM Chondrites*

Supported by petrographic observations, XRD remains the most robust method for identifying heated CM chondrites as it can track mineralogical changes associated with progressive thermal metamorphism such as the dehydration of phyllosilicates to an amorphous phase and recrystallisation of secondary olivine, sulphides, and metal. However, XRD ideally requires some additional supporting characterisation when trying to recognise low level (<300°C) heating that was insufficient to dehydrate phyllosilicates, or meteorites where aqueous alteration overprinted the metamorphic event. Fig. 7a shows that water abundances decrease with metamorphism, but this is not a conclusive characteristic as the water content is related to the relative degree of aqueous alteration prior to dehydration and can also be influenced by terrestrial weathering. The same issue occurs when considering the intensity of the 3 µm feature or hydrogen abundances measured by mass spectrometry (e.g. Alexander et al. 2013). Mid-IR spectra (5 – 25 µm) of Stage II meteorites are indistinguishable from unheated CM chondrites but Stage IIIs are distinct, with spectral features that can be attributed to dehydrated phyllosilicates, olivine, and pyroxene. Features in the mid-IR spectra of Stage IV samples are dominated by olivine and comparable to those of anhydrous carbonaceous meteorites (e.g. CV, CK, CO etc. Beck et al. 2014). Nevertheless, Fig. 7b shows that the 3 µm / 11 µm ratio of heated CMs is correlated with their peak metamorphic temperature.

Thermal metamorphism of CM chondrites is believed to result in bulk oxygen isotopic compositions enriched in $^{17}O$ and $^{18}O$ due to a preferential loss of isotopically "light" water and associated heavy isotope enrichment in the remaining solids (Valley 1986; Clayton and Mayeda 1999, 2009; Lindgren et al. 2020), although shifts to lighter isotopic compositions have been reported for artificial heating experiments (Tonui et al. 2014). Fig. 8 shows that while some of the heated CMs in this study do have heavy bulk oxygen isotopic compositions (e.g. WIS 91600), there is no clear relationship with peak metamorphic temperature. This is not surprising because, like water abundances, bulk oxygen isotopic compositions are likely related to the degree of aqueous alteration prior to metamorphism. In a closed system, reactions between isotopically light anhydrous silicates and heavy fluids results in progressively heavier bulk compositions with increasing hydration (e.g. Clayton and Mayeda 1999), although no clear correlation with the degree of aqueous alteration has been reported. In most cases the bulk oxygen isotopic compositions of heated CM chondrites are still within the range of unheated CMs as most of the oxygen remains in the $SiO_4$ structure of the disordered phase (Nakamura 2006).

With increasing metamorphism, there is a decrease in the abundance of C and N in the heated CM chondrites (Fig. 9). Alexander et al. (2013) showed that at the onset of metamorphism (~Stage II) there is a preferential loss of H from hydrated minerals and organics relative to C, leading to an initial increase in the C/H ratio of heated CMs relative to their unheated counterparts. However, with increasing severity of the metamorphism (Stages III and IV) C is preferentially lost relative to H, possibly due to the breakdown of organics to $CO_2$ and $CH_4$, and the C/H ratio of the most heated samples are significantly lower. The bulk $\delta^{13}C$ values also appear to decrease with metamorphism but there is no systematic change in $\delta D$ or $\delta^{15}N$ (Fig. S6).

Nakamura (2005) proposed that a decrease in the abundance of the volatile trace element Cd was an indicator of thermal metamorphism in CM chondrites. Tonui et al. (2014) reported that separate samples of Murchison artificially heated to between 400°C and 700°C displayed a systematic loss of Cd, with the largest depletions (relative to CI abundances) found for the highest temperatures. Fig. 10 shows that Cd abundances remain fairly constant for the Stage I–III meteorites and then decrease dramatically for the Stage IVs, although the Cd data available for the samples in this study are very limited. The Stage III meteorite on Fig. 10 is EET 83355, which Nakamura (2005) originally classified as a Stage I. However, our Stage III classification is supported by previous elemental abundances, TGA and Raman data that all indicate strong metamorphism (Xiao and Lipschutz 1992; Alexander et al. 2012, 2013; Garenne et al. 2014; Quirco et al. 2018). Furthermore, Mahan et al. (2018) reported depletions in other volatile trace elements for several heated CMs, including EET 83355. Unfortunately, there is little overlap between the samples in this study and Mahan et al. (2018), but they also noted that the most severely metamorphosed CMs (PCA 02010 and PCA 02012) are highly enriched in the heavy isotopes of Zn.

Organic materials are sensitive to thermal metamorphism and Quirico et al. (2018) used a combination of Raman and IR reflectance spectroscopy to characterise IOM in the heated CM chondrites. They reported that the IOM in weakly-to-moderately heated CMs (i.e. Stage II) shows slight structural and chemical modifications that are reflected in narrower D-band widths, higher $CH_2/CH_3$ ratios, and lower aliphatic and carbonyl abundances than unheated CM chondrites. The IOM in more severely heated CMs (i.e. Stage III and Stage IV) has even lower D-band widths, and aliphatic and carbonyl abundances due to the increased structural order and carbonisation of the organics. Again, there is limited overlap in the datasets, but Fig. S7 supports these trends. Using the spectral parameters of Quirico et al. (2018) it is challenging to resolve between Stage III and Stage IV meteorites, although Raman appears to be more sensitive to low levels (<300°C) of thermal metamorphism that cannot be detected from XRD analysis (e.g. QUE 93005).

*5.6 Classification of Heated CM Chondrites*

Heated CM chondrites do not easily fit into the traditional meteorite classification scheme, whereby unaltered samples are type 3, aqueously altered samples are type 1 to 2, and thermally metamorphosed samples are types 4 to 6 (e.g. Weisberg et al. 2006). Furthermore, evidence of thermal metamorphism is not always apparent in polished sections, especially for the Stage II meteorites, unless specifically searched for. A classification scheme for the heated CMs is important though, as an increasing number are discovered in meteorite collections (>36% of the 39 CM and $C2_{ung}$ chondrites studied in Quirico et al. 2018) and the likely occurrence of dehydrated materials in the samples returned from asteroids Ryugu and Bennu (Kitazato et al. 2019, 2021; Hamilton et al. 2019). Tonui et al. (2014) proposed adding the letter "T" (denoting thermal metamorphism) and the metamorphic stages of Nakamura (2005) to a heated CM chondrites' original classification, for example EET 96029 would become a CM2TII. We support this suggestion, acknowledging that XRD remains one of the most effective methods for identifying the heating stage of a CM chondrite. In all likelihood, there will be a continuum both within and between the heating stages, as demonstrated for the Stage IIs that show different degrees of water loss. We do not believe a higher resolution classification scheme using this method is appropriate at this time as it relies on a combination of multiple techniques (XRD, TGA and IR) and more importantly assumes that the conditions (e.g. duration, number/sequence of events) of metamorphism were the same for each meteorite, which is currently unknown.

Finally, we note that except for the Sutter's Mill fall, which contains a close mixture of hydrated and dehydrated phases (Zolensky et al. 2014), all of the heated CM chondrites are finds, most of which were recovered from Antarctica. This maybe reflects sampling of a different population of materials to the contemporary flux of extraterrestrial materials. Ryugu shows evidence of metamorphosed phyllosilicates (Kitazato et al. 2020, 2021) and the upcoming analysis of samples successfully returned from its surface by Hayabusa2, plus those collected from Bennu by OSIRIS-REx, alongside detailed characterisation of heated CM chondrites, will be crucial for understanding the thermal history of hydrous asteroids.

# 6. CONCLUSIONS

There are a number of CM chondrites that following low temperature (<150°C) aqueous alteration also experienced thermal metamorphism at temperatures >300°C. Similarities between the reflectance spectra of heated CM chondrites and the surfaces of low albedo C-type asteroids suggest that these dehydrated materials may be widespread in the solar system. The thermal metamorphism was short-lived, likely caused by impacts and/or solar radiation, but it remains unclear as to how the heated CM chondrites are related to their unheated equivalents. We have used PSD-XRD, TGA and transmission IR spectroscopy to characterise the mineralogy and water contents of 13 CM and one ungrouped carbonaceous chondrite, each of which has previously been assigned a heating stage (I–IV) in the scheme of Nakamura (2005). In summary, we find that:-

(1) Stage I (<300°C) samples show no changes in their mineralogy or water abundances and are indistinguishable from unheated CM chondrites using PSD-XRD, TGA and IR spectroscopy. Low level heating can only be identified using TEM or Raman spectroscopy, while it is also possible that thermal metamorphism in the Stage I meteorites was overprinted by a later period of aqueous alteration. The XRD patterns of Stage II (300–500°C) samples show no coherent diffraction from phyllosilicates due to the loss of structural -OH. Stage II samples contain abundant partially dehydrated amorphous phyllosilicates, which is reflected in water contents (7.5 – 11.0 wt%) and intensity of their 3 µm features being lower than those of the unheated CM chondrites.

(2) Stage III (500–750°C) samples contain both dehydrated amorphous phyllosilicates and secondary olivine and Fe-sulphides that recrystallised during thermal metamorphism. They have complex mid-IR spectra with features of comparable intensity at ~9.9 µm and ~11.2 µm attributed to dehydrated phyllosilicates and olivine, respectively. The water contents (3.4 – 6.3 wt%) and intensity of 3 µm features for Stage III samples are low. The XRD patterns and mid-IR spectra of Stage IV (>750°C) meteorites show that their mineralogy is dominated by secondary olivine, although the water contents (2.9 – 7.4 wt%) and intensity of 3 µm features for the samples in this study are not depleted relative to the Stage IIIs due to terrestrial weathering.

(3) The wide range of peak metamorphic temperatures (~200°C to >750°C) recorded by the heated CM chondrites is consistent with impacts and/or solar radiation being the main source of heat. Variable impact heating could have occurred during the catastrophic disruption of the original parent body or, less likely, at a later stage on a re-accumulated asteroid with a rubble-pile structure. However, some heated CM chondrites appear to lack obvious signs of impact shock implying that recent radiative heating led to dehydration of the asteroid surface. To resolve between the heating mechanisms, we suggest that future petrographic studies systematically search for indicators of shock in the heated CM chondrites.

(4) In addition to mineralogical changes, the relative degree of thermal metamorphism for a heated CM chondrite can be inferred by combining its phyllosilicate/secondary olivine abundance and water content to estimate the overall level of dehydration. We find that the heated CM chondrites have lost ~15 – >65 % of the water they contained at the end of aqueous alteration. If impacts were the main cause of heating, this is consistent with shock pressures of ~20 – 50 GPa.

(5) The heated CM chondrites experienced the same range in degree of aqueous alteration as the unheated CMs, based on their calculated phyllosilicate fractions (PSF) at the onset of thermal metamorphism. There is no clear relationship between the degree of aqueous alteration and peak metamorphic temperatures, which is perhaps not surprising as asteroid surfaces can contain mixtures of materials with different alteration histories.

(6) With increasing heating stage, the C, N and volatile trace element abundances of heated CM chondrites decrease, and the organics are systematically modified (e.g. decreasing FWHM-D band Raman parameter). However, there is no relationship between bulk $\delta^{18}$O values and heating stages because oxygen isotopic compositions are also related to the degree of aqueous alteration prior to metamorphism. We have demonstrated that XRD remains the most robust method for identifying and characterising CM chondrites that were heated to temperatures >300°C, but a higher resolution classification scheme is currently not possible as the exact conditions (e.g. duration, number/sequence of events) of metamorphism remain poorly known and may not have been the same for each meteorite.

**ACKNOWLEDGEMENTS**

We would like to thank the Natural History Museum (NHM), London, and the National Institute of Polar Research (NIPR), Japan, for providing samples analysed in this study. US Antarctic meteorite samples are recovered by the Antarctic Search for Meteorites (ANSMET) program which has been funded by NSF and NASA, and characterised and curated by the Department of Mineral Sciences of the Smithsonian Institution and Astromaterials Curation Office at NASA Johnson Space Center. Jens Najorka and Rachel Norman are thanked for assistance with the XRD and TGA analyses, respectively. The IR spectra were acquired with the help of Gianfelice Cinque, Mark Frogley and Natasha Stephen during experiments SM9614 and SM18213 at Diamond Light Source, UK. Finally, we would like to thank Tomoki Nakamura, Chris Haberle, an anonymous reviewer, and the associate editor Pierre Beck for their insightful comments that significantly improved this manuscript. This work was funded by the Science and Technology Facilities Council (STFC), UK, through grants ST/M00094X/1 and ST/R000727/1, and UK Research and Innovation grant MR/T020261/1.

**Figure 1.** Cartoon showing the mineralogical changes associated with each metamorphic stage in the Nakamura (2005) classification scheme for heated CM chondrites. Stage I meteorites consist of hydrated phyllosilicates and tochilinite, which in Stage II meteorites have been dehydrated to an amorphous phase and poorly crystalline Fe-sulphides by metamorphism at >300°C. Stage III meteorites contain both dehydrated amorphous phyllosilicates, secondary recrystallised silicates and decomposed Fe-sulphides and oxides. Stage IV meteorites are fully dehydrated and contain recrystallised silicates, Fe-sulphides, decomposed carbonates, and metal.

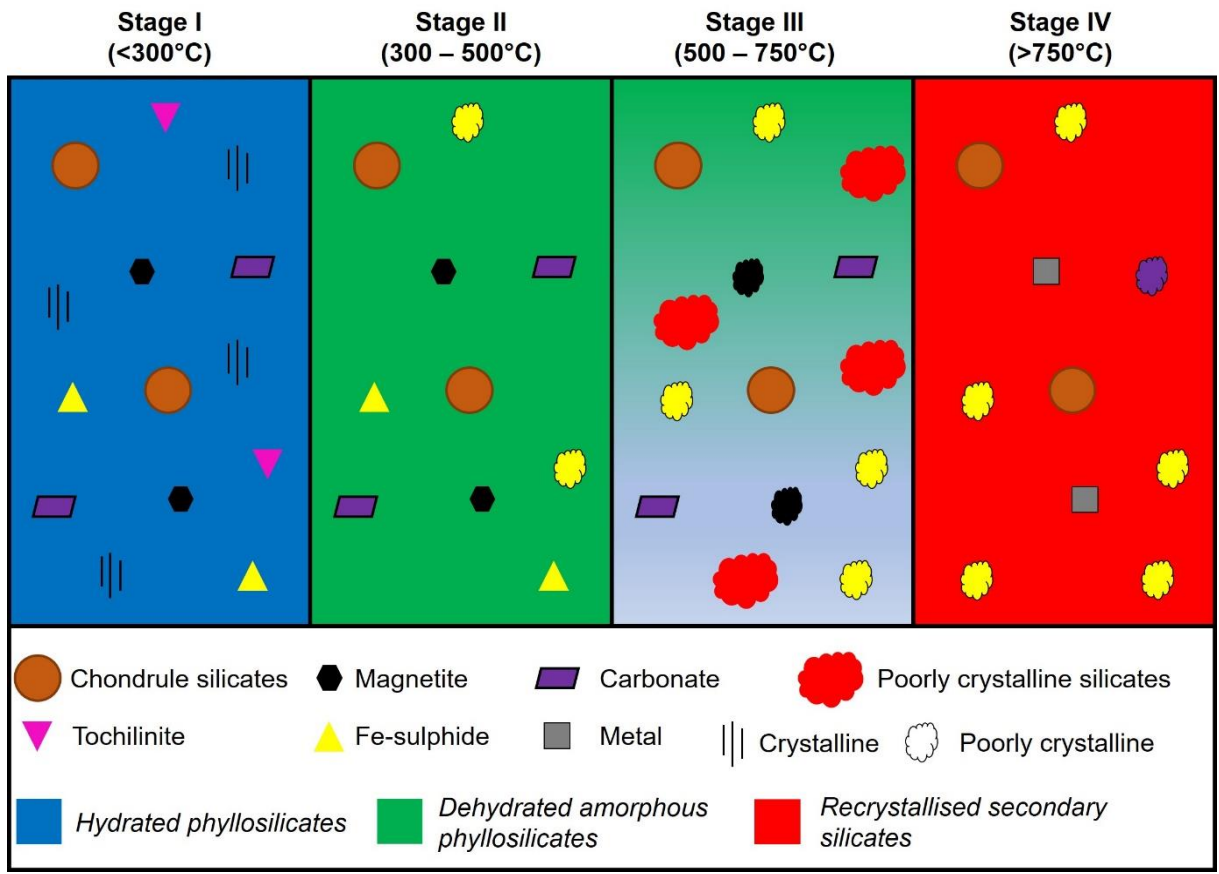

**Figure 2.** Representative PSD-XRD patterns for Stage I to IV heated CM chondrites. The labelled diffraction peaks are Srp = serpentine, Mag = magnetite, Cal = calcite, Ol = olivine, En = enstatite, FeS = Fe-sulphides (troilite and pyrrhotite) and metal. The samples shown are QUE 93005 (Stage I), EET 87522 (Stage II), EET 83355 (Stage III) and PCA 02010 (Stage IV).

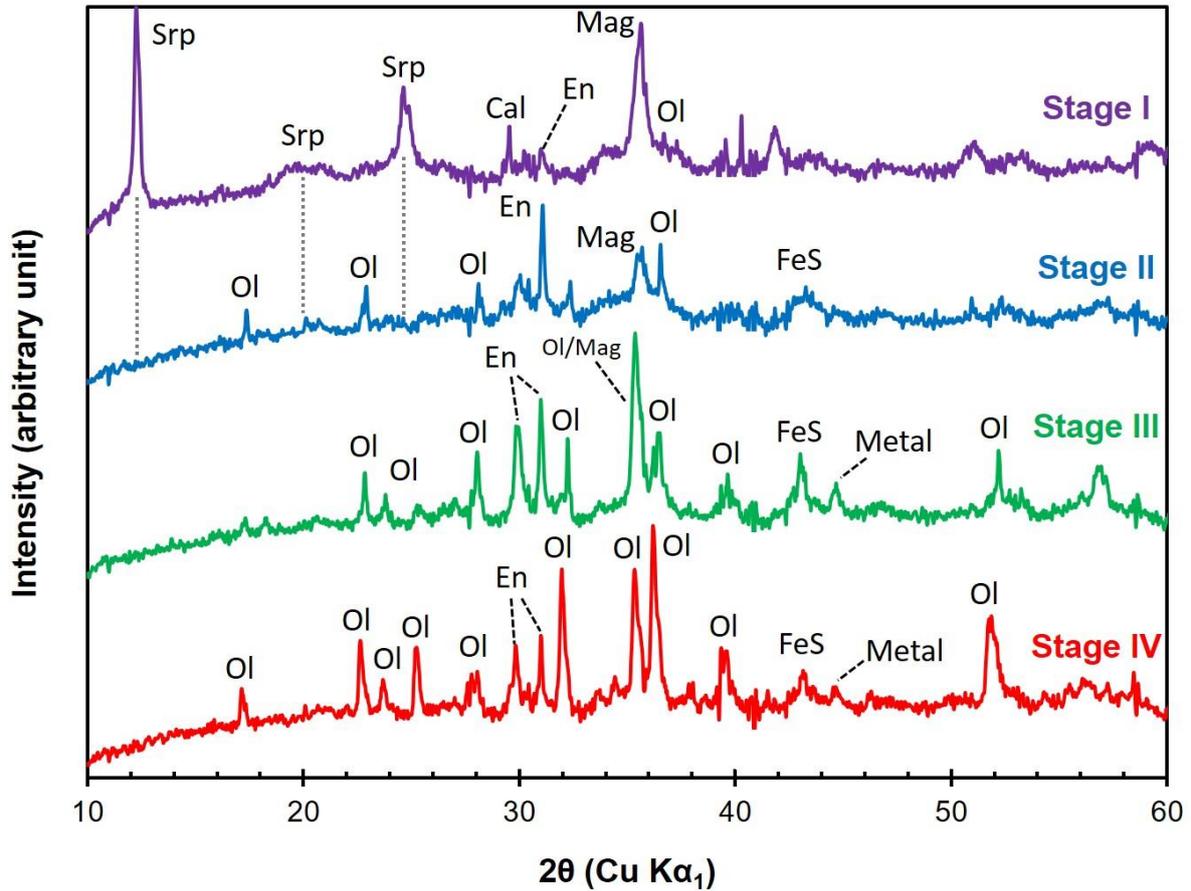

**Figure 3.** Representative mass loss and DTG curves for Stage I to IV heated CM chondrites. Mass loss events are attributed to terrestrial adsorbed water and sulphates (<200°C), the release of -OH/$H_2O$ from Fe-(oxy)hydroxides (200 – 400°C) and phyllosilicates (400 – 770°C), and $CO_2$ from the breakdown of carbonates (770 – 900°C). Mass loss between 200 – 770°C is used to estimate the abundance of water in a meteorite. The samples shown are QUE 93005 (Stage I), EET 87522 (Stage II), EET 83355 (Stage III) and PCA 02010 (Stage IV).

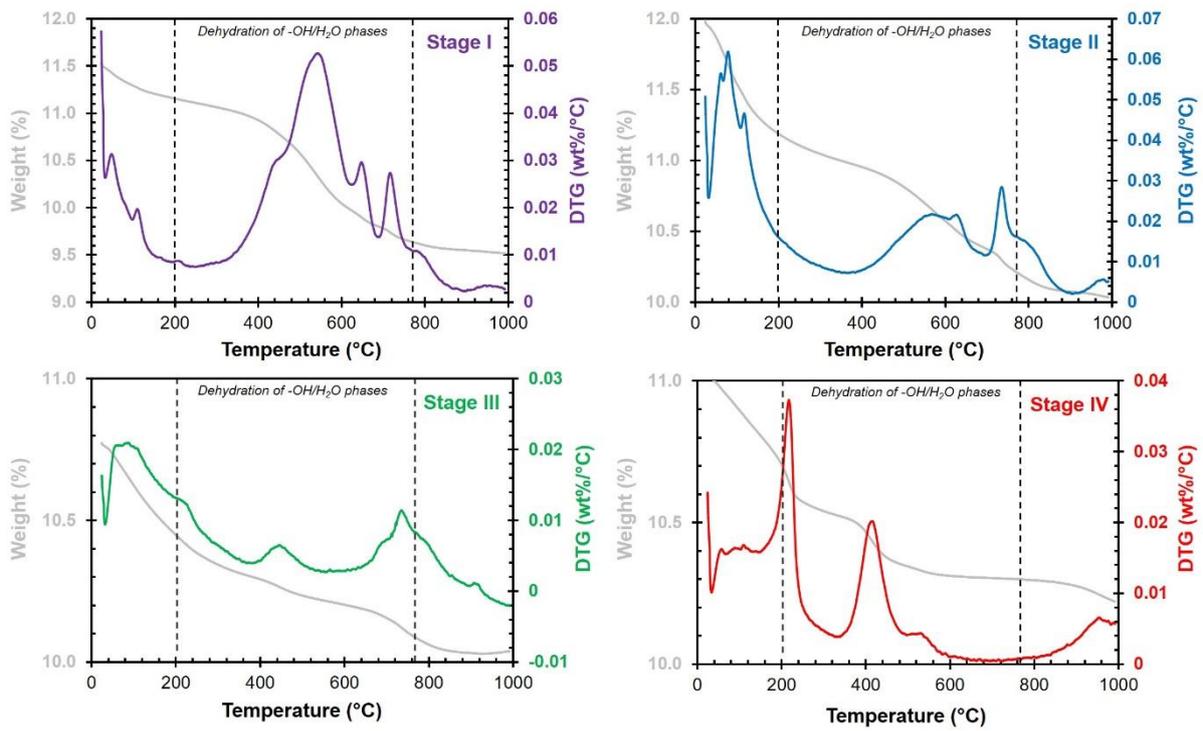

**Figure 4.** Representative transmission mid-IR spectra for Stage I to IV heated CM chondrites. (a) The 5 – 25 μm region, with features attributed to the main Si-O stretching and bending modes of phyllosilicates (~9.9 μm, ~16 μm and ~22 μm), olivine (~11.2 μm) and pyroxene (~9.3 μm). Features at shorter wavelengths are related to carbonates and/or sulphates. (b) The "3 μm" region, with a sharp feature at ~2.72 μm from the -OH stretching mode, and a broader feature extending from 2.7 to 3.2 μm attributed to -OH/$H_2O$ within hydrous phases such as phyllosilicates, sulphates, and rusts.

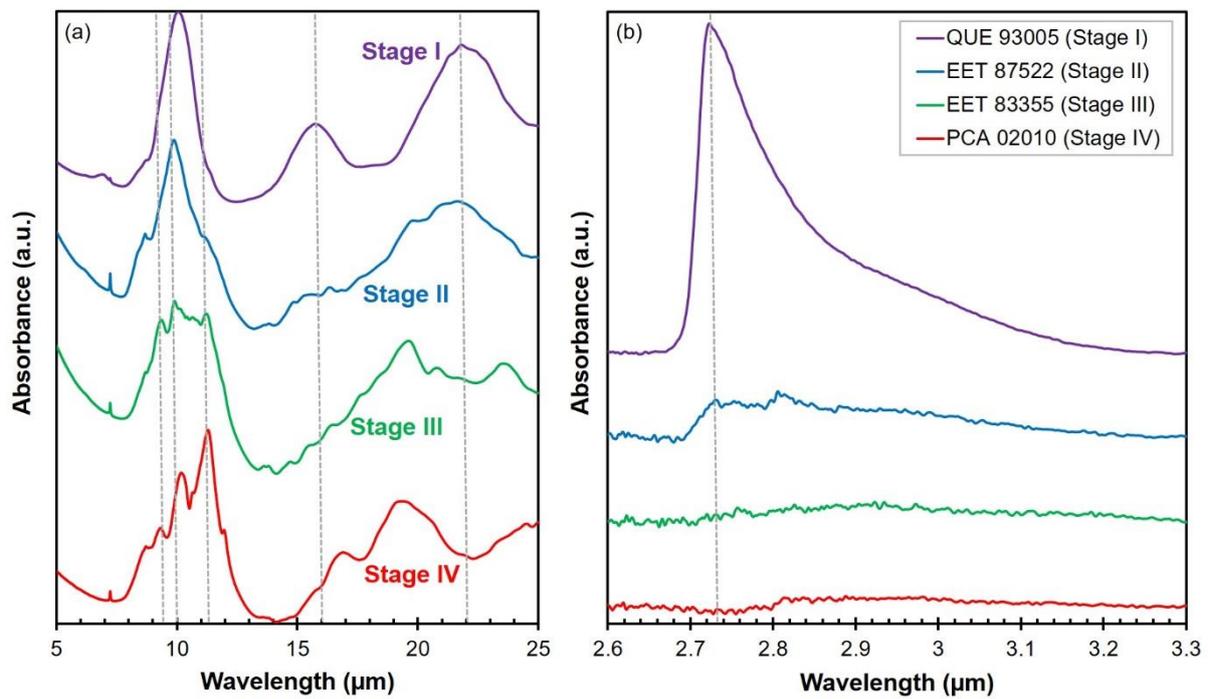

**Figure 5.** Phyllosilicate abundances (both hydrated and dehydrated) plotted against TGA derived water abundances for unheated and heated CM chondrites. The unheated meteorites include Murchison and Santa Cruz measured in this study, and data collected in our laboratory and reported in Mason et al. (2018), Bates et al (2020), and King et al (2020). In unheated CM chondrites aqueous alteration leads to an increase in the abundance of phyllosilicates and water. Thermal metamorphism at 300 – 500°C (Stage II) causes dehydration and a decrease in water content but no change in phyllosilicate abundance. At higher temperatures (>500°C, Stage III and IV), water continues to be lost and phyllosilicate abundances decrease as they are recrystallised as anhydrous silicates. Uncertainties are ~1 vol% for phyllosilicate abundances and ~0.1 % for water abundances.

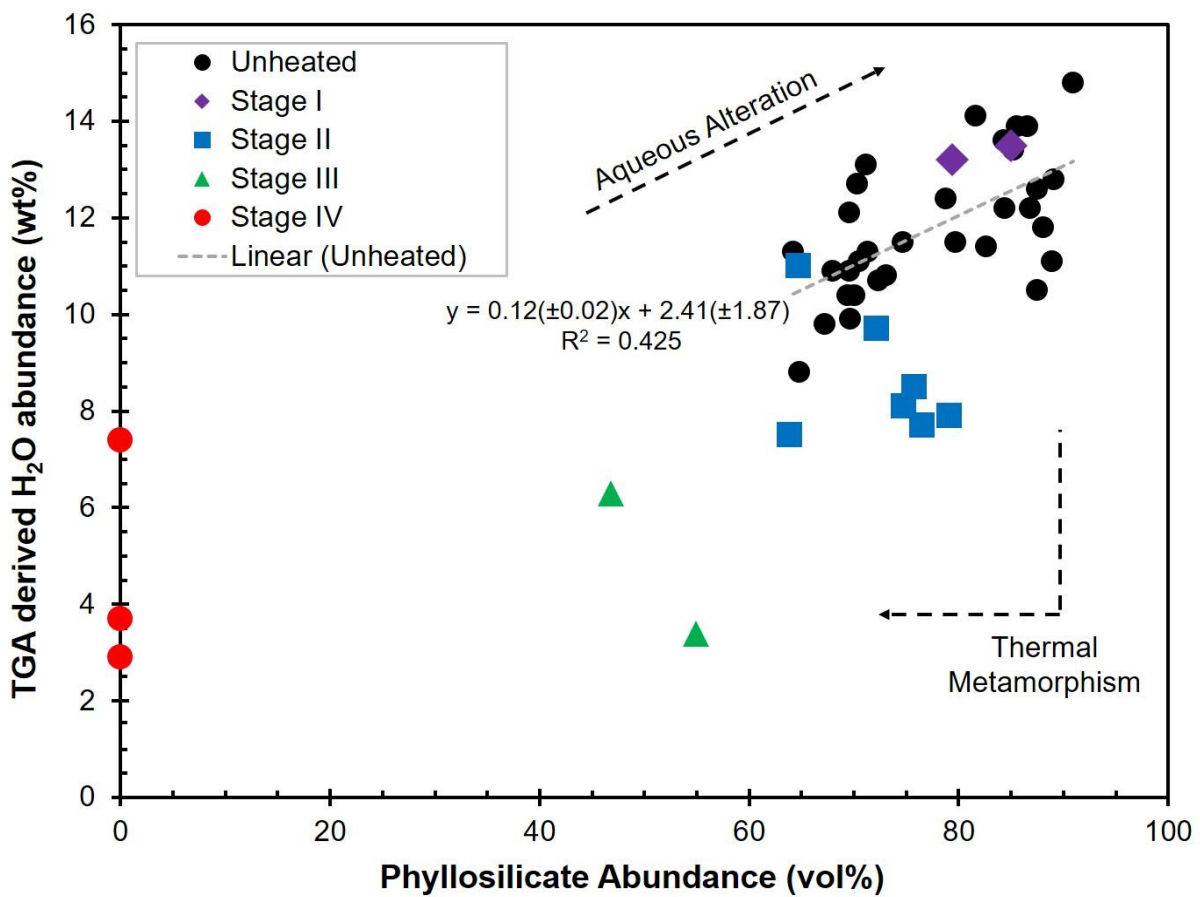

**Figure 6.** Relationship between the degree of aqueous alteration and peak metamorphic temperatures of the heated CM chondrites. (a) Heating stage plotted against the petrologic type based on the phyllosilicate fraction (PSF) at the end of aqueous alteration. (b) The fraction of water lost during thermal metamorphism of Stage II heated CMs plotted against petrologic type at the end of aqueous alteration.

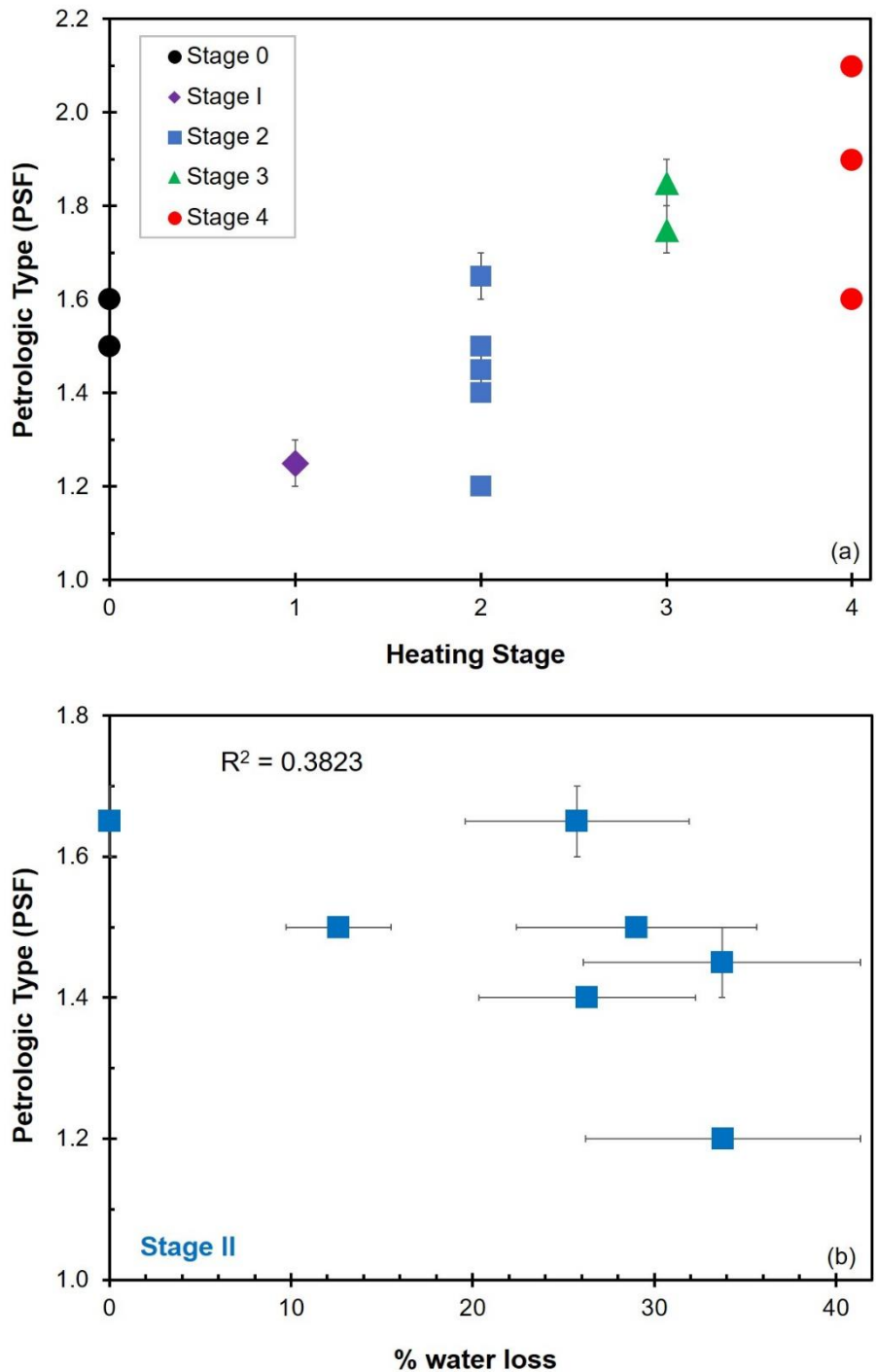

**Figure 7.** (a) TGA derived water abundance (wt%) and (b) 3 µm / 11 µm intensity ratio plotted against heating stage for all CM chondrites analysed in this study.

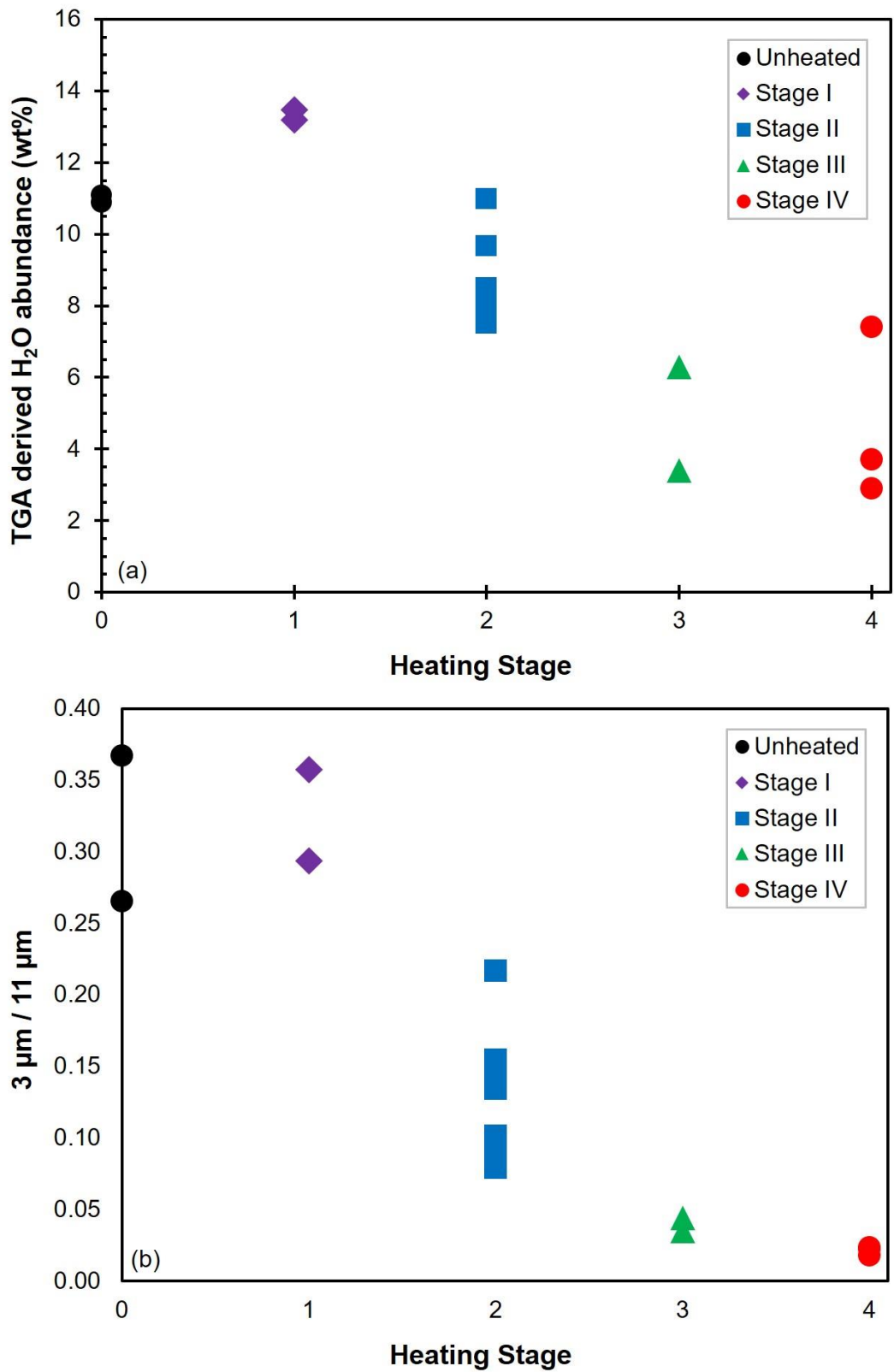

**Figure 8.** Relationship between bulk δ$^{18}$O values and the heating stage of CM chondrites analysed in this study. The range for unheated CMs (i.e. Stage 0) is indicated by the black box. Data are taken from Clayton and Mayeda (1999), Tonui et al. (2014), Ruzicka et al. (2015), and Lee et al. (2016).

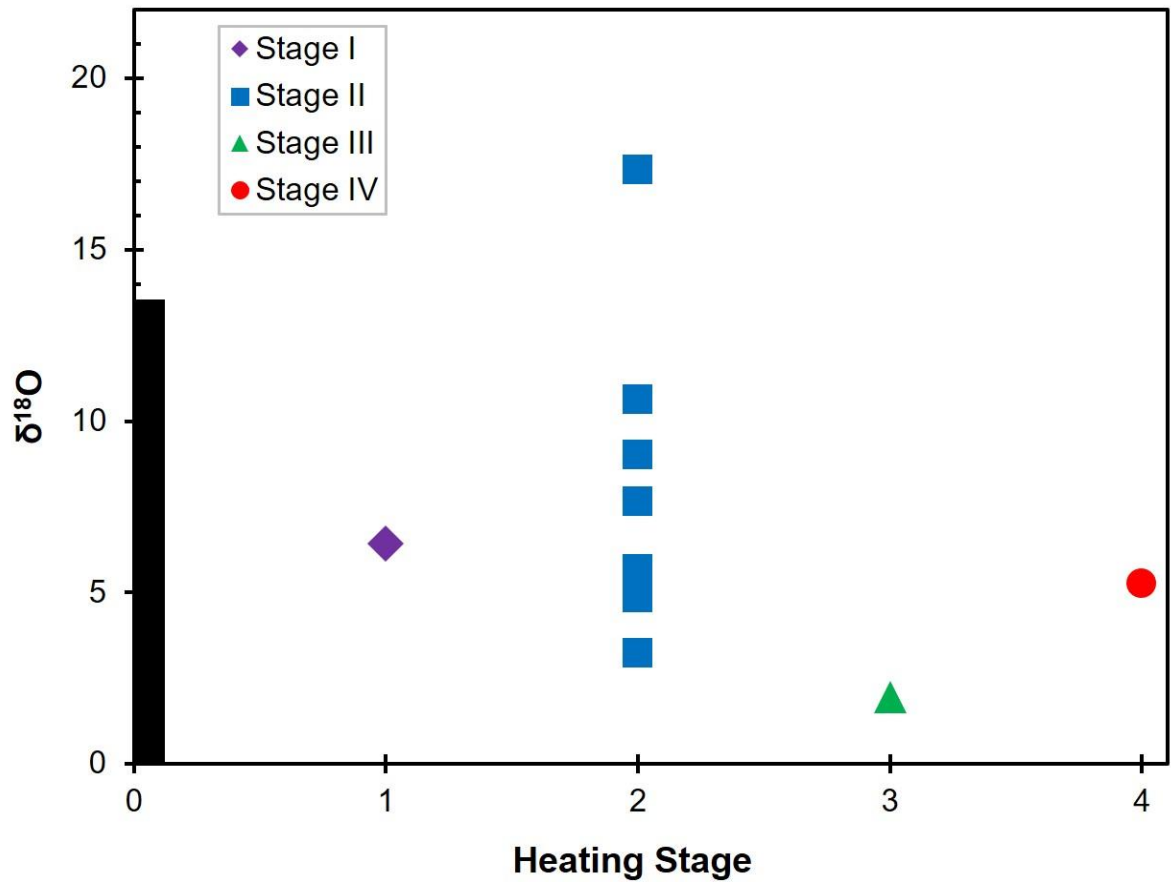

**Figure 9.** Relationship between bulk (a) carbon and (b) nitrogen abundances and the heating stage of CM chondrites analysed in this study. The ranges for unheated CMs (i.e. Stage 0) are indicated by the black boxes. Data are taken from Alexander et al. (2012, 2013), Yabuta et al. (2010), and King et al. (2019a).

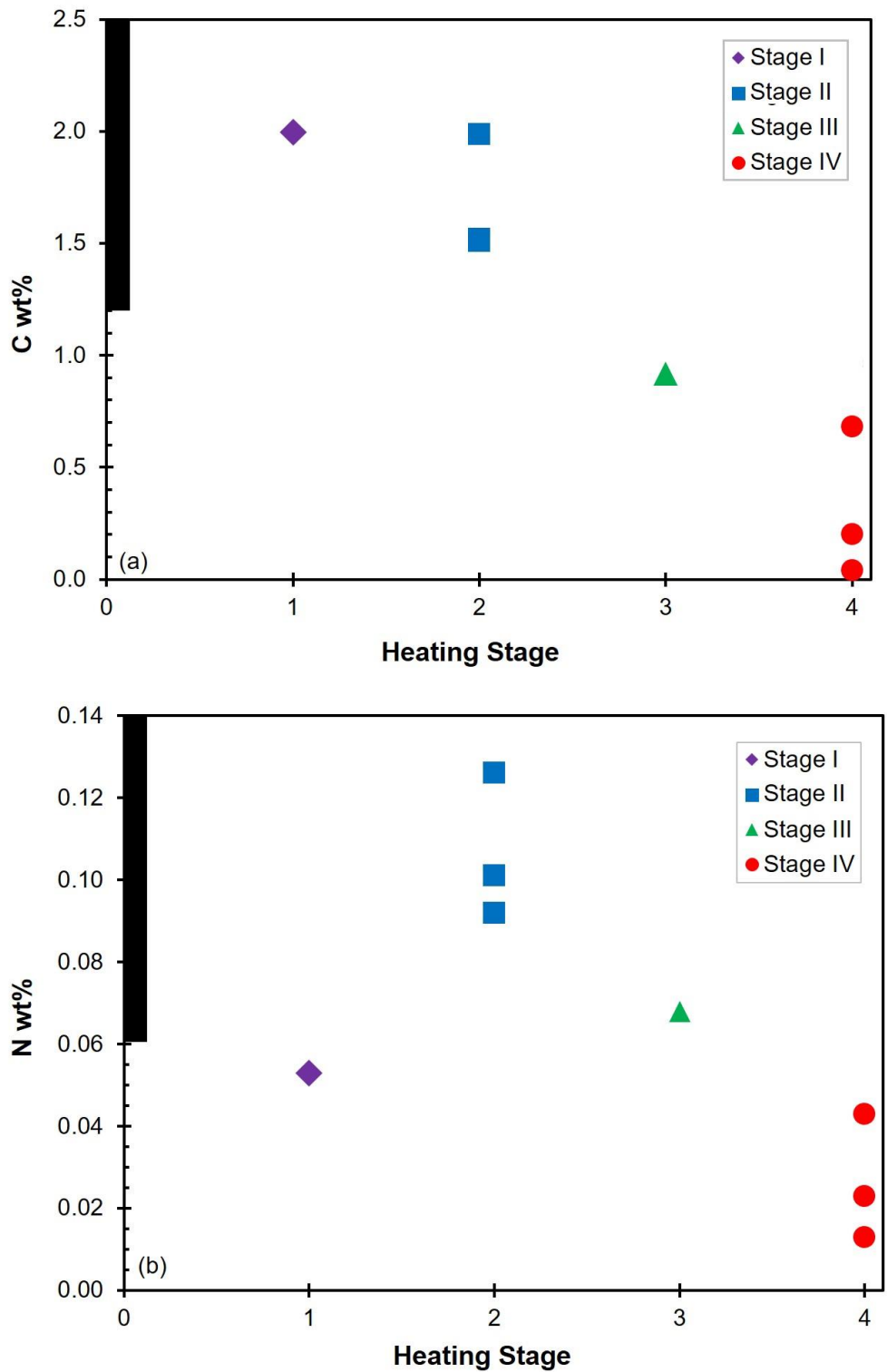

**Figure 10.** Relationship between bulk Cd abundances and the heating stage of CM chondrites analysed in this study. Data are taken from Xiao and Lipschutz (1992), Wang and Lipschutz (1998), and King et al. (2019a).

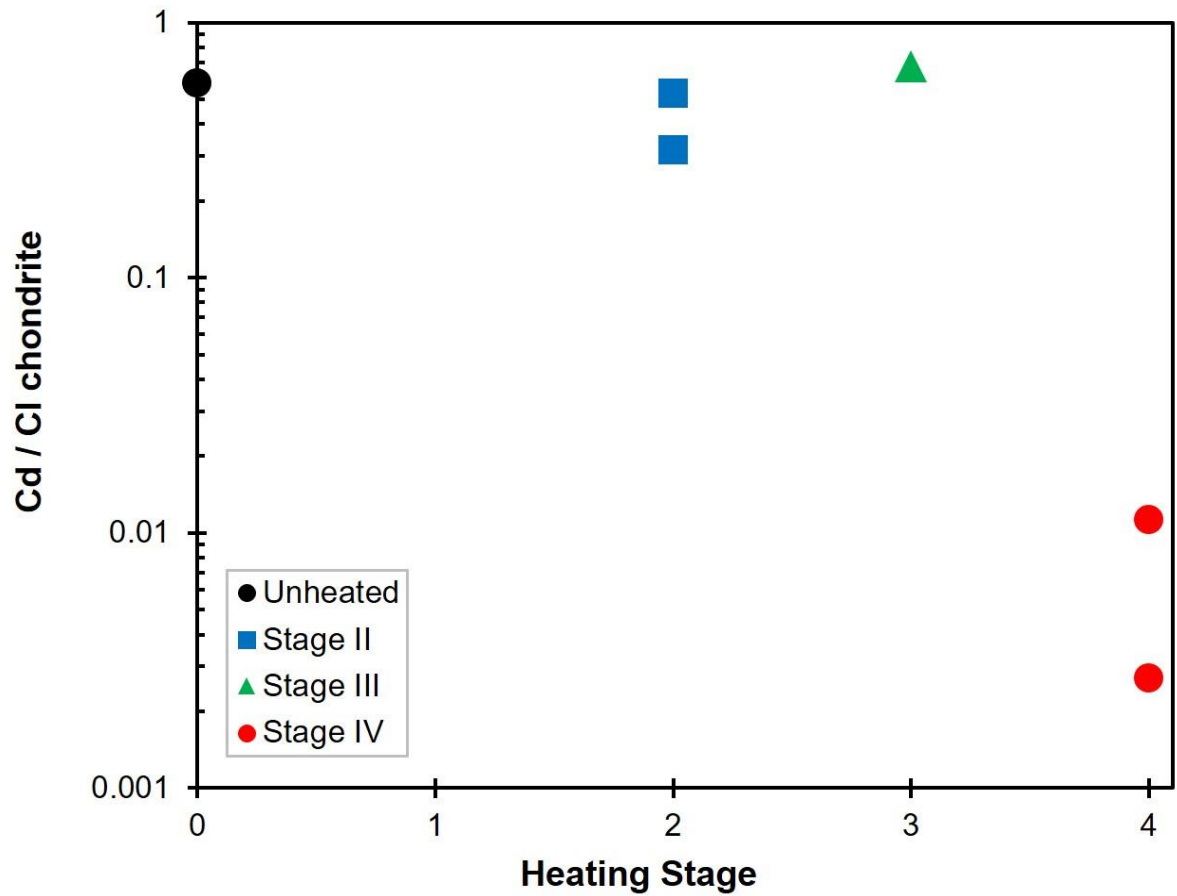

1 **Table 1.** Modal mineralogy of unheated (Stage 0) and heated (Stage I to IV) CM carbonaceous chondrites. The phyllosilicate fraction (PSF) is calculated from
2 the abundance of hydrated, dehydrated amorphous and recrystallised (secondary olivine) phyllosilicates and is used to assign each sample a petrologic type in
3 the Howard et al. (2015) classification scheme. Uncertainties are <0.5 vol% for crystalline phases and <1 vol% for non-crystalline phases, resulting in PSF
4 errors of ~0.01 – 0.02.

| Sample (split number) | Heating Stage* | Heating Stage** | Hydrated Phyllosilicate (vol%) | Amorphous Phyllosilicate (vol%) | Primary Olivine (vol%) | Secondary Olivine (vol%) | Pyroxene (vol%) | Fe-sulphide (vol%) | Magnetite (vol%) | Metal (vol%) | Carbonate (vol%) | Other[†] (vol%) | PSF | Petrologic Type |
|---|---|---|---|---|---|---|---|---|---|---|---|---|---|---|
| Murchison | 0 | 0 | 67.7 | | 19.8 | | 6.9 | 0.6 | 2.1 | | 1.3 | 1.6 | 0.72 | 1.6 |
| Santa Cruz[a] | 0 | 0 | 70.6 | | 16.3 | | 3.9 | 3.9 | 3.0 | | 2.3 | | 0.78 | 1.5 |
| A-881458 (*70*) | I | I | 84.8 | | 9.3 | | 1.6 | 0.5 | 2.2 | 0.2 | 1.0 | 0.4 | 0.89 | 1.3-1.2 |
| QUE 93005 (*34,7*) | I | I | 79.4 | | 8.7 | | 1.9 | 3.9 | 5.0 | | 1.1 | | 0.88 | 1.3-1.2 |
| EET 87522 (*51,2*) | II | I/II | | 74.8 | 15.3 | | 6.6 | 0.7 | 1.9 | 0.1 | 0.4 | 0.2 | 0.77 | 1.5 |
| EET 96029 (*57,21*)[b] | II | II | | 64.7 | 12.2 | | 15.7 | 2.3 | 3.1 | 0.3 | 0.9 | 0.8 | 0.70 | 1.7-1.6 |
| Jbilet Winselwan[a] | II | II | | 72.2 | 16.6 | | 5.1 | 3.7 | 1.1 | | | 1.3 | 0.77 | 1.5 |
| Y-793321 (*114*) | II | II | | 63.9 | 22.7 | | 6.3 | 1.8 | 2.6 | 0.2 | 2.5 | | 0.69 | 1.7-1.6 |
| WIS 91600 (*76,9*) | II | II | | 79.1 | 5.3 | | 1.5 | 3.9 | 9.1 | | 1.1 | | 0.92 | 1.2 |
| Y-86695 (*103*) | II | II/III | | 75.8 | 12.2 | | 5.2 | 3.5 | 2.2 | 0.2 | 0.9 | | 0.81 | 1.4 |
| Y-82098 (*108*) | II | III/IV | | 76.5 | 13.7 | | 5.1 | 1.1 | 2.4 | 0.2 | 1.0 | | 0.80 | 1.5-1.4 |
| EET 83355 (*33,0*) | III | I/II | | 54.9 | 13.8 | 3.2 | 18.8 | 2.1 | 6.3 | 0.5 | 0.4 | | 0.64 | 1.8-1.7 |
| Y-82054 (*103*) | III | III | | 46.8 | 19.6 | 4.6 | 16.6 | 6.0 | 4.0 | 0.5 | | 1.9 | 0.59 | 1.9-1.8 |
| PCA 91008 (*37,5*)[c] | IV | III | | | 22.7 | 54.0 | 16.9 | 0.8 | 2.5 | | | 3.1 | 0.58 | 1.9 |
| PCA 02010 (*24,2*)[c] | IV | IV | | | 9.5 | 65.7 | 16.1 | 3.8 | | 0.4 | 1.3 | 3.2 | 0.72 | 1.6 |
| PCA 02012 (*21,3*) | IV | IV | | | 28.0 | 38.6 | 15.8 | 13.4 | | | 0.8 | 3.4 | 0.47 | 2.1 |

5 Abundances previously reported in (a) King et al. (2019a) (Jbilet Winselwan sample no. BM2013, M4), (b) Lee et al. (2016), and (c) Hanna et al. (2020).
6 *Heating stage based on XRD patterns in this study. **Heating stage based on literature data.
7 [†]Includes sulphates, rusts and feldspar.

**Table 2.** Mass loss (wt%) as a function of temperature measured by TGA and intensity of the 3 µm and 11 µm features in IR spectra for unheated (Stage 0) and heated (Stage I to IV) CM carbonaceous chondrites.

| Sample (*split number*) | Heating Stage* | Heating Stage** | 25 – 200°C | 200 – 400°C | 400 – 770°C | 770 – 900°C | $H_2O$ (200 – 770°C) | 3 µm feature | 11 um feature | 3 / 11 µm ratio |
|---|---|---|---|---|---|---|---|---|---|---|
| Murchison | 0 | 0 | 4.3 | 4.0 | 7.1 | 1.2 | 11.1 | 0.069 | 0.188 | 0.367 |
| Santa Cruz | 0 | 0 | 3.0 | 4.1 | 7.0 | 0.9 | 11.1 | 0.075 | 0.283 | 0.265 |
| A-881458 (*70*) | I | I | 4.9 | 3.7 | 9.8 | 1.0 | 13.5 | 0.080 | 0.224 | 0.357 |
| QUE 93005 (*34,7*) | I | I | 3.0 | 2.0 | 11.2 | 0.8 | 13.2 | 0.059 | 0.201 | 0.294 |
| EET 87522 (*51,2*) | II | I/II | 6.5 | 1.9 | 6.2 | 1.2 | 8.1 | 0.054 | 0.367 | 0.147 |
| EET 96029 (*57,21*)[a] | II | II | 8.6 | 2.3 | 8.7 | 1.8 | 11.0 | 0.054 | 0.350 | 0.154 |
| Jbilet Winselwan[b] | II | II | 4.5 | 2.2 | 7.5 | 1.5 | 9.7 | 0.055 | 0.410 | 0.134 |
| Y-793321 (*114*) | II | II | 4.8 | 1.9 | 5.6 | 1.1 | 7.5 | 0.033 | 0.418 | 0.079 |
| WIS 91600 (*76,9*) | II | II | 5.0 | 0.5 | 7.4 | 2.0 | 7.9 | 0.060 | 0.277 | 0.217 |
| Y-86695 (*103*) | II | II/III | 5.6 | 1.9 | 6.6 | 1.0 | 8.5 | 0.035 | 0.385 | 0.091 |
| Y-82098 (*108*) | II | III/IV | 4.5 | 2.0 | 5.7 | 1.0 | 7.7 | 0.038 | 0.374 | 0.102 |
| EET 83355 (*33,0*) | III | I/II | 3.0 | 1.4 | 1.9 | 0.5 | 3.3 | 0.026 | 0.742 | 0.035 |
| Y-82054 (*103*) | III | III | 5.3 | 2.0 | 4.3 | 1.0 | 6.3 | 0.022 | 0.496 | 0.044 |
| PCA 91008 (*37,5*) | IV | III | 6.6 | 2.3 | 5.1 | 0.5 | 7.4 | 0.026 | 1.115 | 0.023 |
| PCA 02010 (*24,2*) | IV | IV | 2.9 | 2.2 | 1.5 | 0.2 | 3.7 | 0.019 | 1.073 | 0.018 |
| PCA 02012 (*21,3*) | IV | IV | 4.0 | 0.8 | 2.1 | 0.5 | 2.9 | 0.024 | 1.062 | 0.023 |
| | | | | | | | | | | |
| CM chondrites[c] | 0 | | 1.0 – 9.8 | 1.0 – 4.7 | 5.0 – 12.9 | 0.0 – 2.3 | 6.2 – 15.2 | | | |

TGA data previously reported in (a) Lee et al. (2016) and (b) King et al. (2019a) (Jbilet Winselwan sample no. BM2013, M4).
(c) Unheated CM chondrite data from Garenne et al. (2014), Mason et al. (2018), Bates et al. (2020), and King et al. (2020).
*Heating stage based on XRD patterns in this study. **Heating stage based on literature data.

**Figure S1.** PSD-XRD pattern for WIS 91600 shown alongside a model pattern. The model pattern is constructed by summing together the mineral standard patterns in their relative proportions (value in parentheses) as determined during the profile-stripping routine. Subtracting the model from the meteorite pattern results in a residual of zero counts, indicating that all major and minor phases are accounted for. Diffraction peaks for Fe- and Mg-serpentines are not detected in the heated CMs and as such are not used in the profile-stripping but can still produce large negative peaks in the residual.

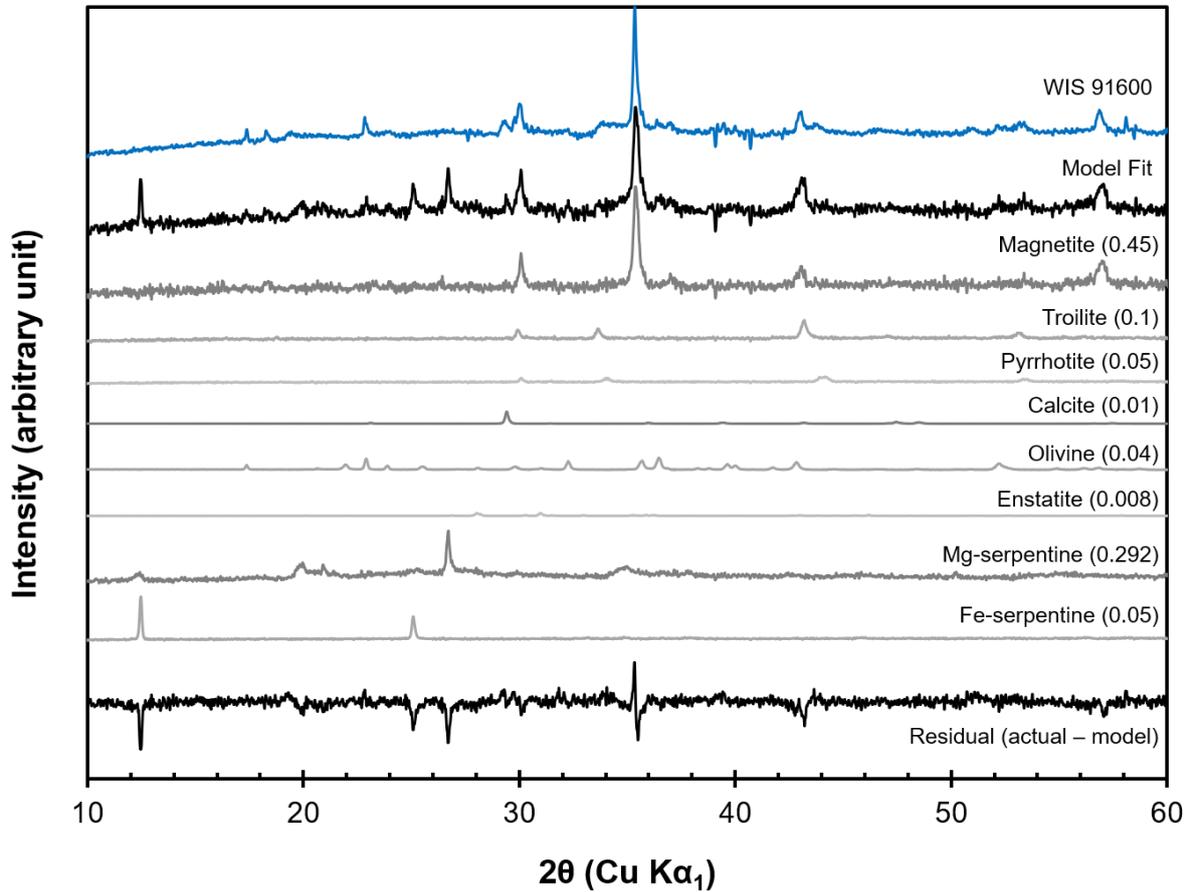

**Figure S2.** Transmission mid-IR spectra for saponite (BM1938, 1225), antigorite (BM1922, 908), olivine (San Carlos) and enstatite (BM1928, 285) mineral standards taken from the Natural History Museum, London, collection. The main features are attributed to Si-O stretching and bending modes (~9.3 µm ~9.9 µm, ~11.2 µm, ~16 µm and ~22 µm).

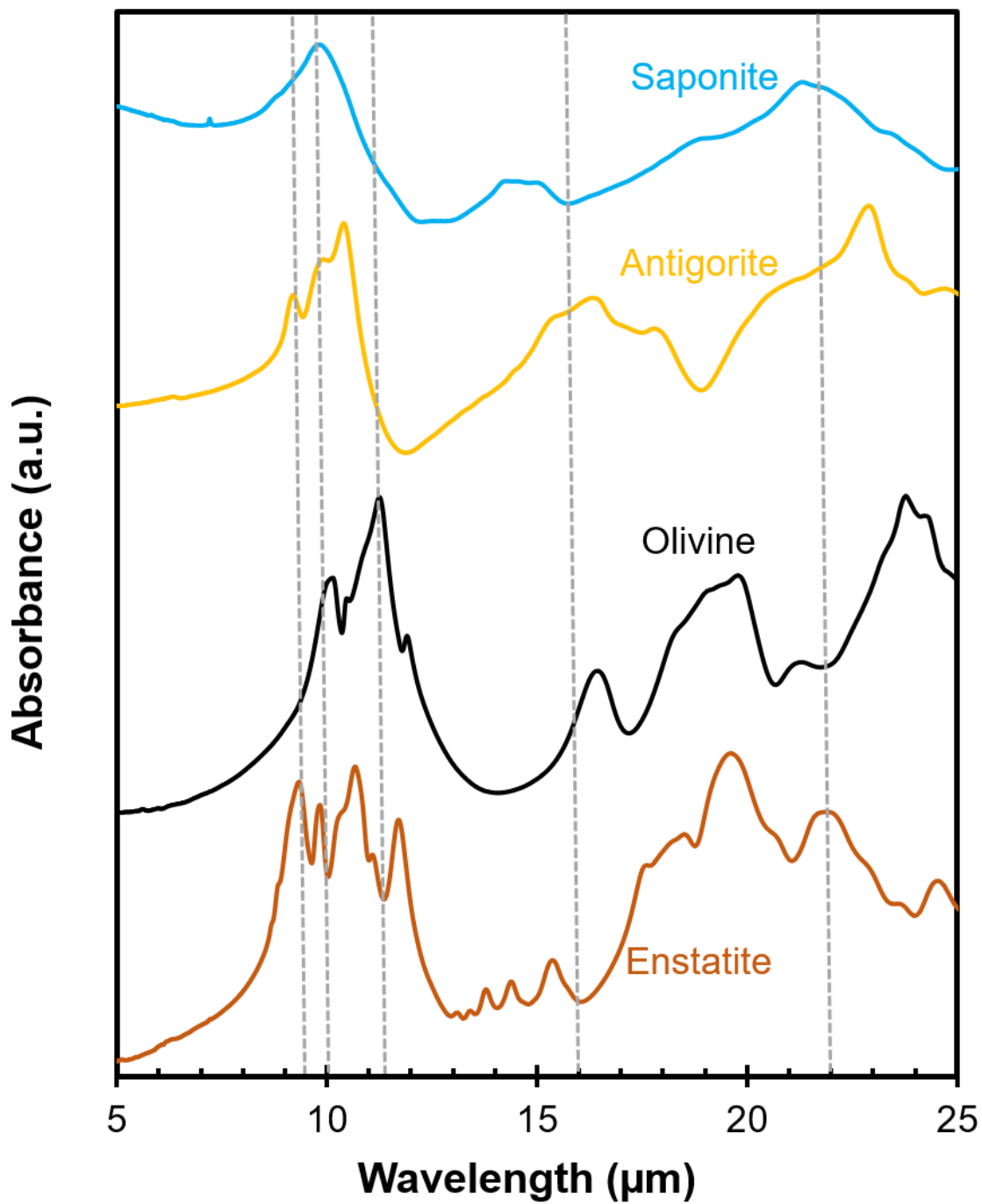

**Figure S3.** Transmission mid-IR spectra of the heated CM chondrites Jbilet Winselwan, WIS 91600 (CM$_{an}$) and PCA 91008 (CM2$_{an}$). Jbilet Winselwan differs from other CM chondrites as it contains multiple intense peaks (at ~8.5 µm, ~8.9 µm, ~14.8 µm and ~16.8 µm) from sulphates; unlike other Stage II meteorites WIS 91600 retains a relatively prominent feature at ~16 µm, possibly from -OH; PCA 91008 has an intense peak at ~8.5 µm from sulphates.

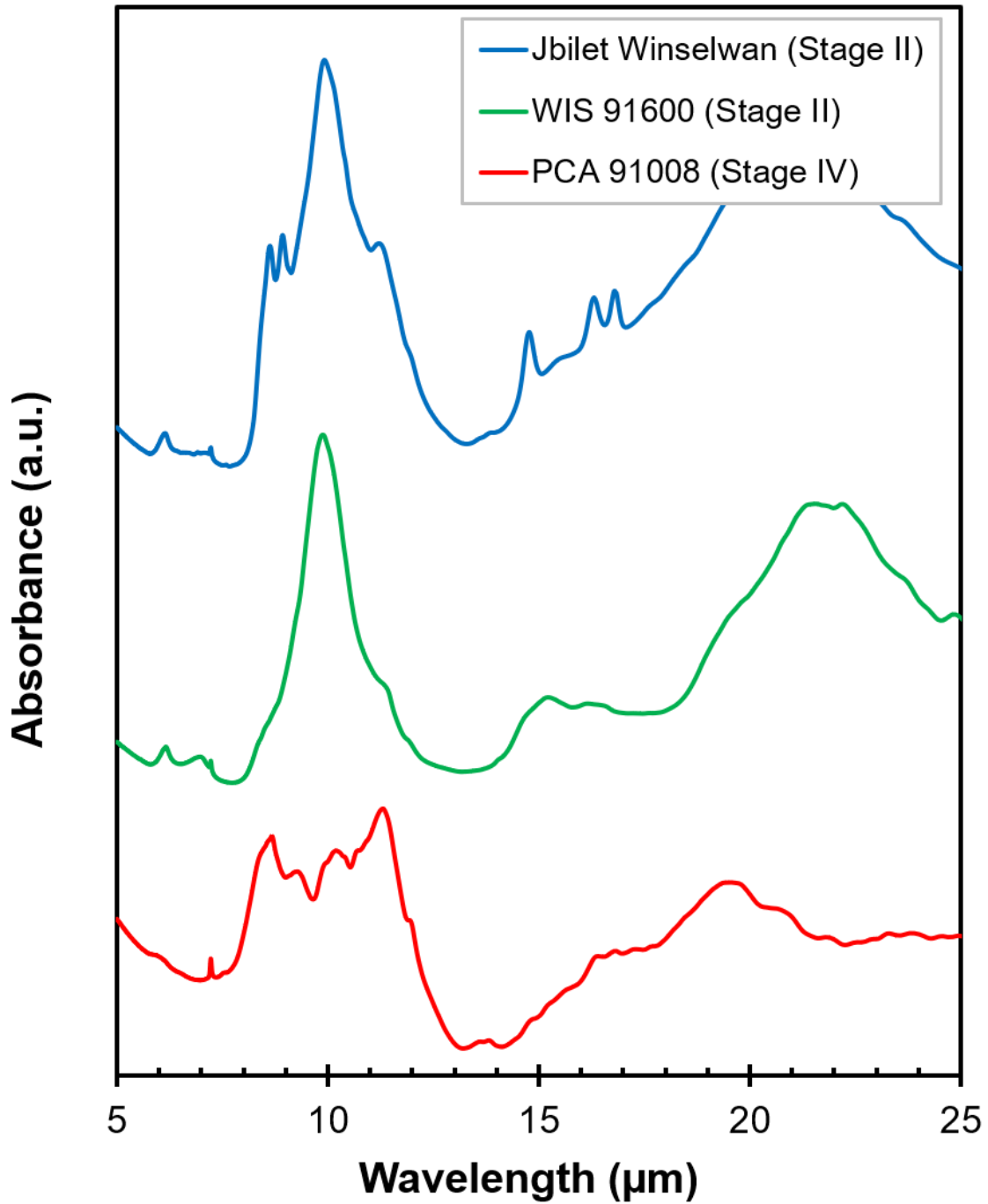

**Figure S4.** Correlation between the (a) intensity of the 3 μm feature and TGA derived water contents and (b) intensity of the 11 μm feature and olivine abundances from PSD-XRD for unheated and heated CM chondrites.

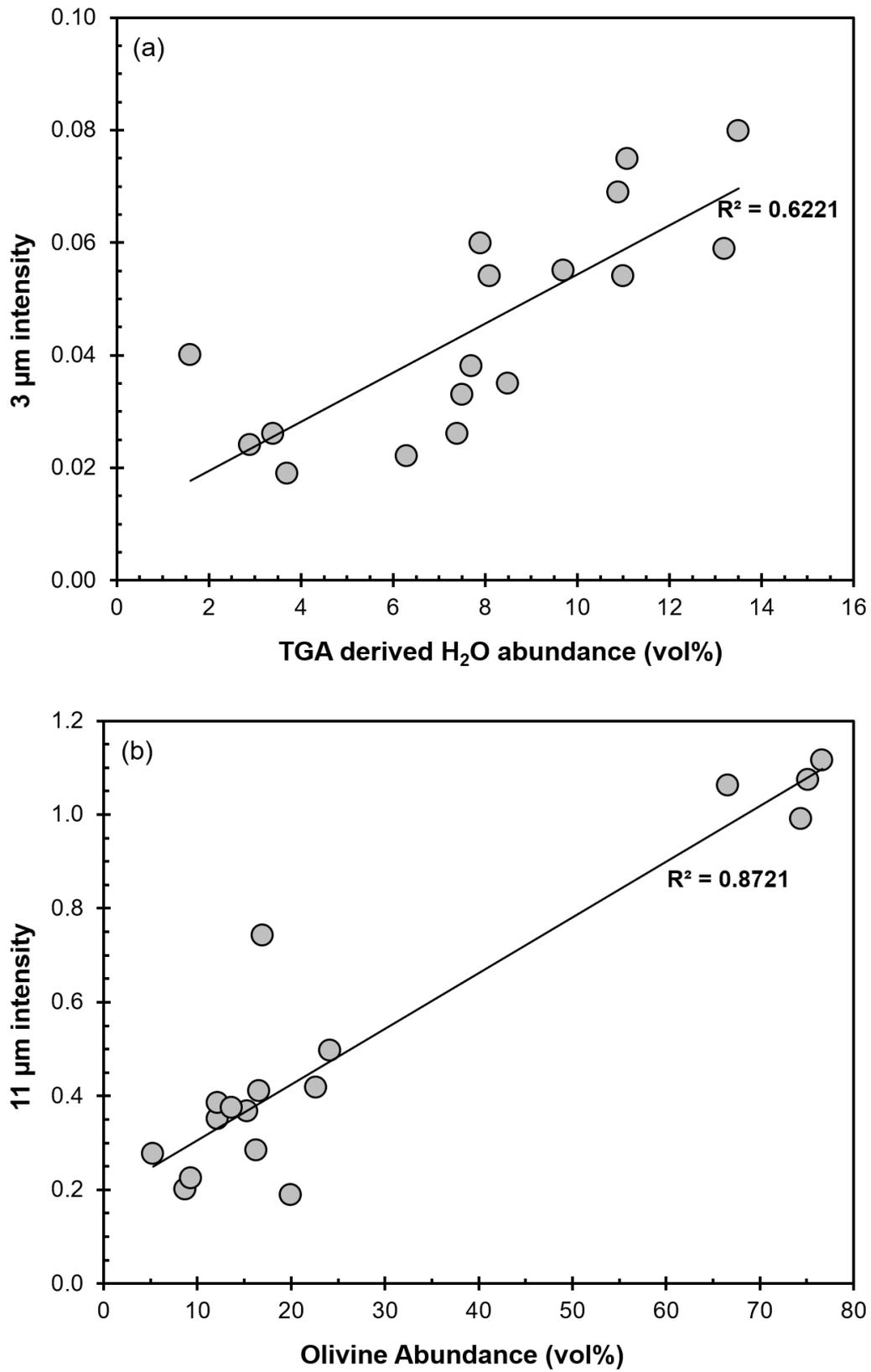

**Figure S5.** Phyllosilicate abundances (both hydrated and dehydrated) plotted against the intensity of the 3 μm feature for unheated and heated CM chondrites.

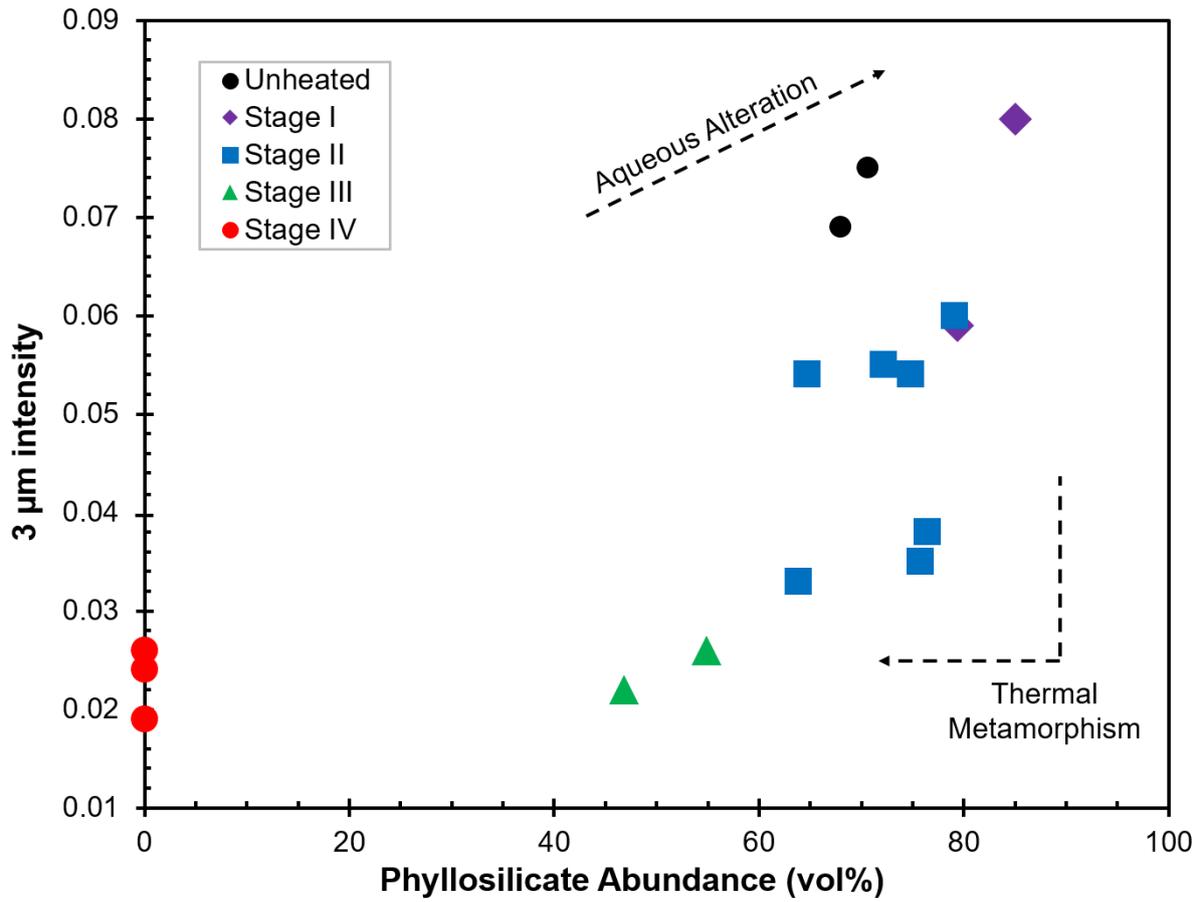

**Figure S6.** Relationship between bulk (a) $\delta^{13}C$ (b) $\delta D$ and (c) $\delta^{15}N$ values and the heating stages of CM chondrites analysed in this study. Typical ranges for unheated CMs (i.e. Stage 0) are indicated by the black boxes. Data are taken from Alexander et al. (2012, 2013), Yabuta et al. (2010), and King et al. (2019a).

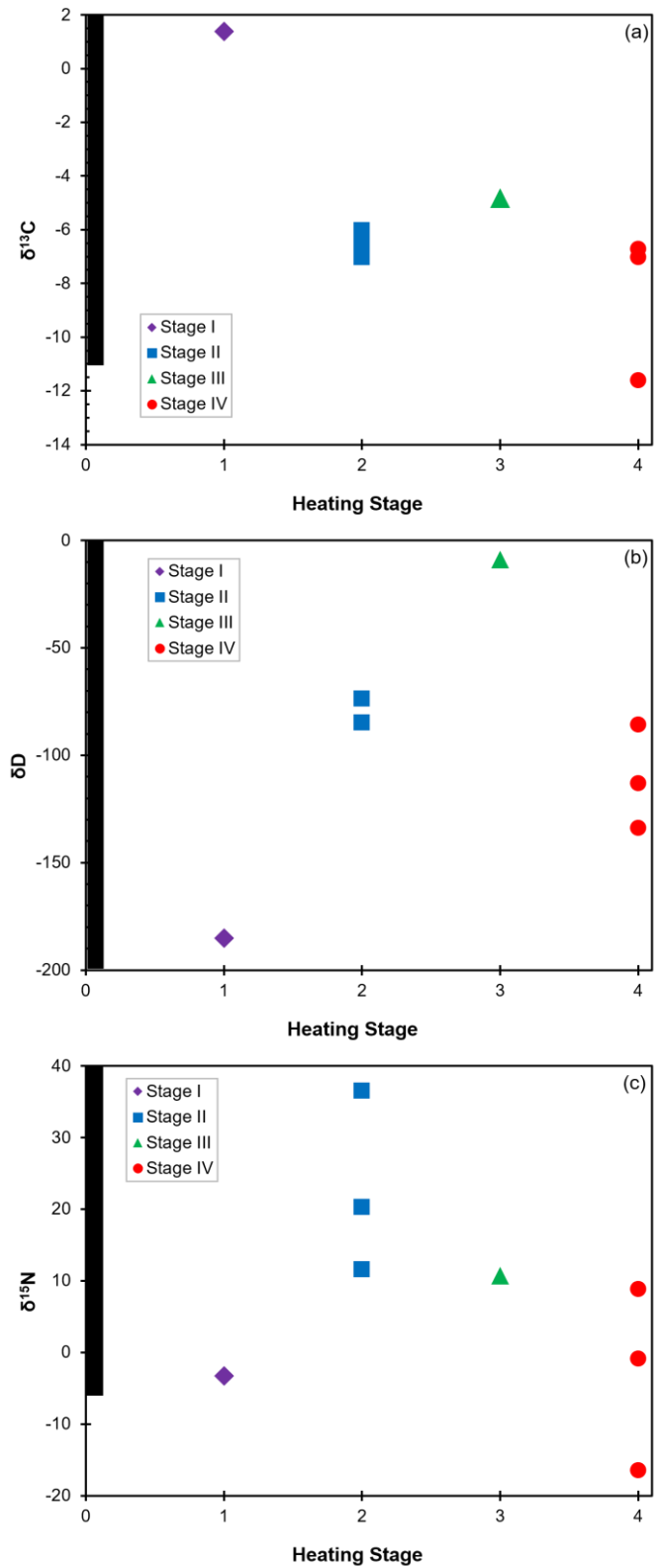

**Figure S7.** Relationship between the (a) FWHM-D band Raman parameter (b) $CH_2/CH_3$ ratio and (c) $CH+CH_2+CH_3$ of the IOM and the heating stages of CM chondrites analysed in this study. Data are taken from Quirico et al. (2018) and the unheated (Stage 0) sample is Murchison.

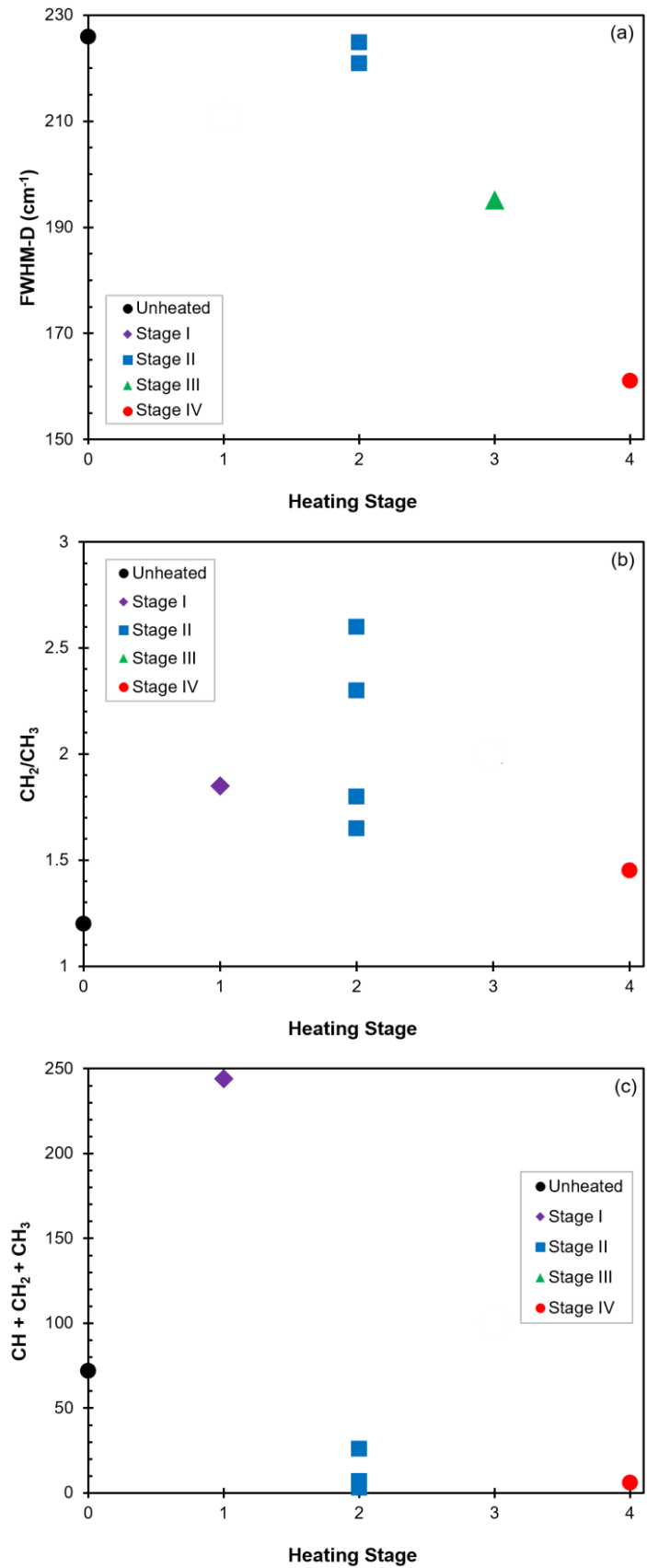

**Table S1.** Mass loss (wt%) as a function of temperature measured by TGA in five separate aliquots of the Murchison CM2 chondrite.

| Aliquot | 25 – 200°C | 200 – 400°C | 400 – 770°C | 770 – 900°C | $H_2O$ (200 – 770°C) |
|---|---|---|---|---|---|
| Mur-1 | 3.9 | 4.0 | 7.1 | 1.1 | 11.1 |
| Mur-2 | 4.5 | 4.3 | 6.8 | 1.2 | 11.1 |
| Mur-3 | 4.2 | 3.7 | 7.1 | 1.2 | 10.8 |
| Mur-4 | 4.2 | 3.7 | 7.6 | 1.3 | 11.3 |
| Mur-5 | 4.9 | 4.3 | 7.0 | 1.2 | 11.3 |
| **average** | **4.3** | **4.0** | **7.1** | **1.2** | **11.1** |
| *(stdev)* | *(0.4)* | *(0.3)* | *(0.3)* | *(0.1)* | *(0.2)* |
| Garenne et al. (2014) | 2.7 | 3.5 | 7.3 | 1.9 | 10.8 |